\documentclass[aps,superscriptaddress,showpacs,prd,twocolumn]{revtex4-2}
\usepackage{amsfonts}
\usepackage{amssymb}
\usepackage{amsmath}
\usepackage{color}
\usepackage[usenames,dvipsnames]{xcolor}
\usepackage{float}
\usepackage{accents}
\usepackage{graphicx}
\usepackage{epstopdf}
\usepackage{ulem}
\usepackage[colorlinks=true,pdfstartview=FitV,linkcolor=blue,citecolor=blue,urlcolor=blue,breaklinks=true]{hyperref}
\usepackage{eurosym}
\usepackage{amssymb}
\usepackage{amsfonts}
\usepackage{amsmath}
\usepackage{graphicx}
\usepackage{bm}
\usepackage{color}
\usepackage[usenames,dvipsnames]{xcolor}
\usepackage[dvips]{epsfig}
\usepackage[dvips]{graphicx}
\usepackage{float}
\usepackage{epstopdf}
\usepackage[colorlinks=true,pdfstartview=FitV,linkcolor=blue,citecolor=blue,urlcolor=blue,breaklinks=true]{hyperref}
\usepackage{wasysym}
\usepackage{blindtext}
\usepackage{longtable}
\usepackage{dcolumn}

\begin{document}

\title{{BPS solitons with internal structures in a
restricted baby Skyrme-Maxwell theory in a magnetic medium}}
\author{J. Andrade}
\email{joao.luis@discente.ufma.br}
\author{R.\ Casana}
\email{rodolfo.casana@ufma.br}\affiliation{Departamento de F\'{\i}sica, Universidade Federal do Maranh\~{a}o,
65080-805, S\~{a}o Lu\'{\i}s, Maranh\~{a}o, Brazil.}
\author{E. da Hora}
\email{carlos.hora@ufma.br}\affiliation{Coordena\c{c}\~{a}o do Curso de Bacharelado Interdisciplinar em Ci\^{e}ncia e Tecnologia,\\
Universidade Federal do Maranh\~{a}o, {65080-805}, S\~{a}o Lu\'{\i}s, Maranh\~{a}o, Brazil.}
\author{{Andr\'e C.} Santos}
\email{andre.cs@discente.ufma.br}\affiliation{Departamento de F\'{\i}sica, Universidade Federal do Maranh\~{a}o,
65080-805, S\~{a}o Lu\'{\i}s, Maranh\~{a}o, Brazil.}

\begin{abstract}
We consider {a {restricted baby Skyrme-Maxwell scenario} e%
}nlarged via the inclusion of a nontrivial magnetic permeability. {%
We then proceed with the {{minimization} of its total}
energy by means of the Bogomol'nyi-Prasad-Sommerfield (BPS) prescription,
from which we get that the self-dual potential now depends on the magnetic
permeability itself. {As a result}, we obtain not only the lower
bound for the energy, but also the self-dual equations whose solutions
saturate that bound. In such a context, we focus our attention on those
time-independent gauged skyrmions with radial symmetry and no electric
charge.} {We solve the {effective self-dual} equations
numerically {for} different choices of magnetic permeability, {
from which we obtain} {BPS}\ magnetic fields {whose} internal
structures {form} concentric rings. We also explain analytically the
formation of these structures based on the values of {a single real}
parameter {which characterizes} the respective magnetic {
permeabilities}.}
\end{abstract}

\pacs{11.10.Kk, 11.10.Lm, 11.27.+d}
\maketitle

\section{Introduction}

\label{Intro}

Topologically nontrivial structures are commonly described by means of
those\ time-independent solutions which come from highly nonlinear
Euler-Lagrange equations {\cite{n5}}. In such a context, the potential term
which defines the vacuum manifold of the respective theory not only
introduces the nonlinearity itself, but it is also expected to allow the
spontaneous symmetry breaking mechanism to occur (whose effects include the
formation of a topological profile as a result of the corresponding phase
transition). {The point is that highly nonlinear equations of
motion are typically quite hard to solve. However, this issue can be
circumvented via the minimization of the system's total energy by employing
the Bogomol'nyi-Prasad-Sommerfield (BPS) prescription \cite{n4}. The
implementation of {such} algorithm determines a specific expression
for the potential, {but it also provides} }a {lower bound 
{for the} energy (the BPS bound) and the corresponding BPS equations
whose solutions saturate that bound (and therefore describe energetically
stable configurations). In addition, {it is always possible to verify
that} the BPS equations are compatible with the Euler-Lagrange equations, 
{from which one concludes that} the BPS profiles stand for legitimate
solutions {of} the model. In the literature, there are alternative
methods for the obtainment of such BPS configurations; see, for instance,
the study of the conservation of the energy-momentum tensor \cite{ano}, the
on-shell procedure \cite{onshell}, and the strong-necessary conditions
technique \cite{lukasz}.}

{The} full Skyrme model was proposed in 1961 as a generalized
nonlinear sigma theory defined in $(3+1)$-dimensions {\cite{1a}}. Its
Lagrange density contains the so-called Skyrme term ({a quartic
kinetic, i.e.} a term of degree four in the first-derivative of the scalar
sector), the $\sigma $ term (a quadratic kinetic one), and a potential ({a nonderivative term}) which was originally proposed in order to
study the pion mass. {The Skyrme theory {can be
interpreted as} an effective low-energy model of Quantum Chromodynamics 
{which engenders} stable solitonic structures, so-called skyrmions, 
{which} can be applied {to study} some physical properties of
those hadrons and nuclei {\cite{2a}}.}

In this context, the study of the planar version of the Skyrme theory, known
as the {baby Skyrme model \cite{3a}}, serves to the comprehension of many
aspects of the original $(3+1)$-dimensional scenario, including the
conditions under which it eventually accepts the implementation of the BPS
prescription. {The baby Skyrme model in the absence of the $%
\sigma $-term, named the restricted baby Skyrme model \cite{13a}, supports a
BPS structure \cite{14a}.} Furthermore, over the last years to investigate
other physical phenomena, the skyrmions have also been used to describe
topological quantum Hall effect {\cite{4a}}, in chiral nematic liquid
crystals \cite{5a}, superconductors {\cite{6a}}, brane cosmology {\cite{7a}}%
, magnetic materials {\cite{8a}}, for instance.

Moreover, in order to investigate the electromagnetic properties of the baby
Skyrme model, it is necessary to couple it to {an Abelian gauge
field \cite{16a}}. {In such a context, the BPS skyrmions {
appear in a} restricted baby Skyrme{-Maxwell} model \cite{a1}, 
{and also occur} when {the Skyrme sector is }minimally coupled
to the Chern-Simons term \cite{a2} and to the Maxwell-Chern-Simons action 
\cite{a3}. Additional results on the study of those BPS solutions in a
Skyrme-Born-Infeld scenario {can be found in} \cite{a4}, while
supersymmetric extensions of these restricted gauged baby Skyrme {
theories} are in the Refs. \cite{se, se1, se2, se3,se4}.}

We now go a little bit further into this issue and consider how the
electromagnetic properties of a material medium affect the first-order
skyrmions which arise from a BPS restricted baby Maxwell-Skyrme mode. Here,
these properties are studied via the introduction of a nonstandard function
which multiples the Maxwell term and therefore represents the magnetic
permeability of the medium.

In order to present our results, this manuscript is organised as {
follows. In the Section II}, we introduce the {restricted baby }
Maxwell-Skyrme model enlarged via the inclusion of a nontrivial {
magnetic permeability}. We also present the definitions and conventions
which we adopt in our work. In the sequence, we focus our attention on those
radially symmetric time-independent profiles which describe gauged skyrmions
in a planar context. We then look for first-order solutions via the
minimization of the effective total energy by means of the BPS prescription,
from which it arises a differential constraint whose solution is the
expression for the potential which supports the existence of well-behaved
first-order configurations. As a result of our construction, we obtain not
only the BPS bound itself, but also the first-order BPS equations which
saturate it. Then, in the Sec. III, we use the first-order expressions
obtained previously to define effective BPS scenarios and their
corresponding gauged skyrmions. The point is that both the potential and the
first-order equations depend on the expression for the {magnetic
permeability} explicitly. We then choose such expression in order to
generate BPS skyrmions with internal structures, i.e. which behave
standardly near the boundaries, but exhibit a nonusual profile in the
intermediate regions. In particular, we investigate in detail how the shape
of the magnetic field depends on a free real parameter which enters the
expression for the {magnetic permeability}. We generalize these
results by choosing a {magnetic permeability} which gives rise to BPS
gauged skyrmions with a much more sophisticated internal structure. In this
case, the description of the magnetic profile additionally requires the
application of numerical techniques. However, even in this more intricate
case, our predictions fit the solutions extremely well. Finally, the Sec. IV
brings our ending comments and perspectives regarding future contributions.

In this manuscript, we adopt the {natural units system} and $\eta
^{\mu \nu }=(+--)$ for the metric signature, for the sake of simplicity.

{\color{blue} }

\section{The restricted gauged baby {Skyrme in a} magnetic medium: The BPS
structure \label{2}}

\label{general}

We begin by presenting the $(2+1)$-dimensional restricted gauged baby Skyrme
model enlarged via the inclusion of {{an {a priori}
arbitrary function which represents a nontrivial} magnetic permeability, the
corresponding Lagrangian function reading%
\begin{equation}
L=E_{0}\int d^{2}\mathbf{x}\,{\mathcal{L}}\text{,}  \label{L0}
\end{equation}%
where the factor $E_{0}$ sets the energy scale {of the model} (which
will be taken {as} $E_{0}=1$ hereafter). The Lagrangian density is} 
\begin{equation}
\mathcal{L}=-\frac{G}{4g^{2}}F_{\mu \nu }F^{\mu \nu }-\frac{\lambda ^{2}}{4}%
(D_{\mu }\vec{\varphi}\times D_{\nu }\vec{\varphi})^{2}-V\text{,}  \label{01}
\end{equation}%
where the first term stands for {Maxwell's }{action} now
multiplied by a {magnetic permeability} function $G=G\left( \varphi
_{n}\right) $ which depends on the quantity $\varphi _{n}=\widehat{n}\cdot 
\vec{\varphi}$. In the internal space, $\widehat{n}$ represents an unitary
vector which defines a preferred direction, while the Skyrme field $\vec{%
\varphi}=\left( \varphi _{1},\varphi _{2},\varphi _{3}\right) $ is given as
a triplet of real scalar fields constrained to satisfy $\vec{\varphi}\cdot 
\vec{\varphi}=1$ and\ therefore describing a spherical surface with unitary
radius. Moreover, $F_{\mu \nu }=\partial _{\mu }A_{\nu }-\partial _{\nu
}A_{\mu }$ is the electromagnetic field strength tensor and%
\begin{equation}
D_{\mu }\vec{\varphi}=\partial _{\mu }\vec{\varphi}+A_{\mu }\widehat{n}%
\times \vec{\varphi}
\end{equation}%
stands for the usual covariant derivative of the Skyrme field. The third
term brings the self-interacting potential $V=V\left( \varphi _{n}\right) $,
while both $\lambda $ and $g$ are coupling constants inherent to the model {(which we assume to be {nonnegative} from
now on)}{. Moreover, the Skyrme field and the function $G$ are
dimensionless, {while}}{\ the gauge field, the electromagnetic
constant }$g${\ and the Skyrme one }$\lambda ${\ have mass
dimensions equal to }$1${, }$1${\ and }$-1${,
respectively.}

{It is instructive to write down the Euler-Lagrange equation
for the gauge sector which comes from (\ref{01}), i.e. 
\begin{equation}
\partial _{\nu }\left( GF^{\nu \mu }\right) =g^{2}j^{\mu }\text{,}
\label{02}
\end{equation}%
where $j^{\mu }=\widehat{n}\cdot J^{\mu }$ is the conserved current density,
with%
\begin{equation}
J^{\mu }=\lambda ^{2}\left[ \vec{\varphi}\cdot \left( D^{\mu }\vec{\varphi}%
\times D^{\nu }\vec{\varphi}\right) \right] D_{\nu }\vec{\varphi}\text{.}
\label{03}
\end{equation}%
}

{The Gauss law for time-independent configurations }{reads}
\begin{equation}
\partial _{i}\left( G\partial ^{i}A^{0}\right) =g^{2}j^{0}\text{,}
\label{gl}
\end{equation}%
where%
\begin{equation}
j^{0}=\lambda ^{2}A_{0}\left( \widehat{n}\cdot \partial ^{i}\vec{\varphi}%
\right) \left( \widehat{n}\cdot \partial _{i}\vec{\varphi}\right)
\label{ecd}
\end{equation}%
is the electric charge density. The point is that $A^{0}=0$ stands for a
legitimate gauge choice, given that it solves the Gauss law (\ref{gl})
identically. Thus, we conclude that the stationary configurations we study
in this manuscript are electrically neutral (i.e. present no electric field
and electric charge).

Instead of studying the solutions of the second-order Euler-Lagrange
equations, we focus {our attetion} on those first-order
configurations {which {minimize} the {total
energy of the theory}.} We achieve such a goal via the implementation of the
BPS procedure, {the starting point being the stationary energy
density, which is related to the energy-momentum tensor of the model (\ref%
{01}), i.e.} {
\begin{eqnarray}
T_{\alpha \beta } &=&-\frac{G}{g^{2}}F_{\mu \alpha }F^{\mu }{}_{\beta
}-\lambda ^{2}(D_{\mu }\vec{\varphi}\times D_{\alpha }\vec{\varphi})\cdot
(D^{\mu }\vec{\varphi}\times D_{\beta }\vec{\varphi})  \notag \\[0.2cm]
&&-\eta _{\alpha \beta }\mathcal{L}\text{.}
\end{eqnarray}%
\begin{figure}[tbp]
\includegraphics[width=8.4cm]{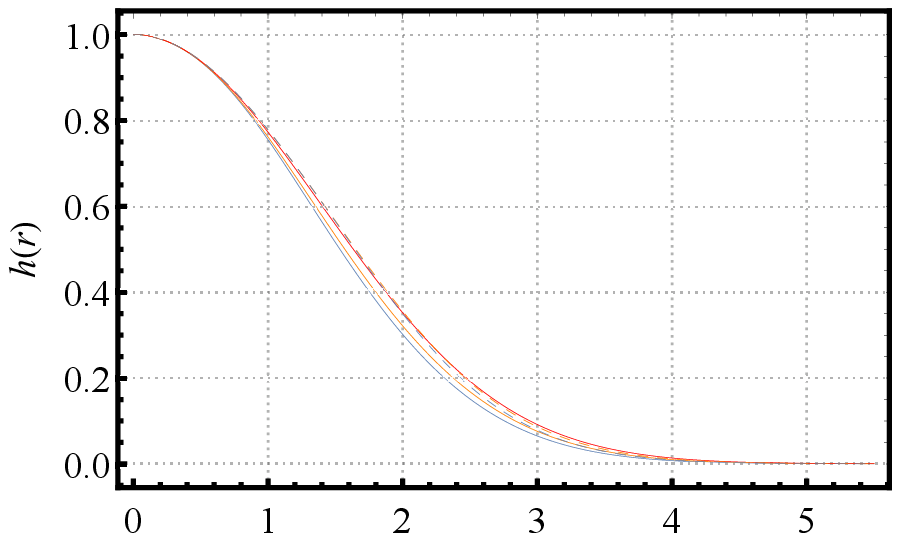}\vspace{0.5cm} %
\includegraphics[width=8.4cm]{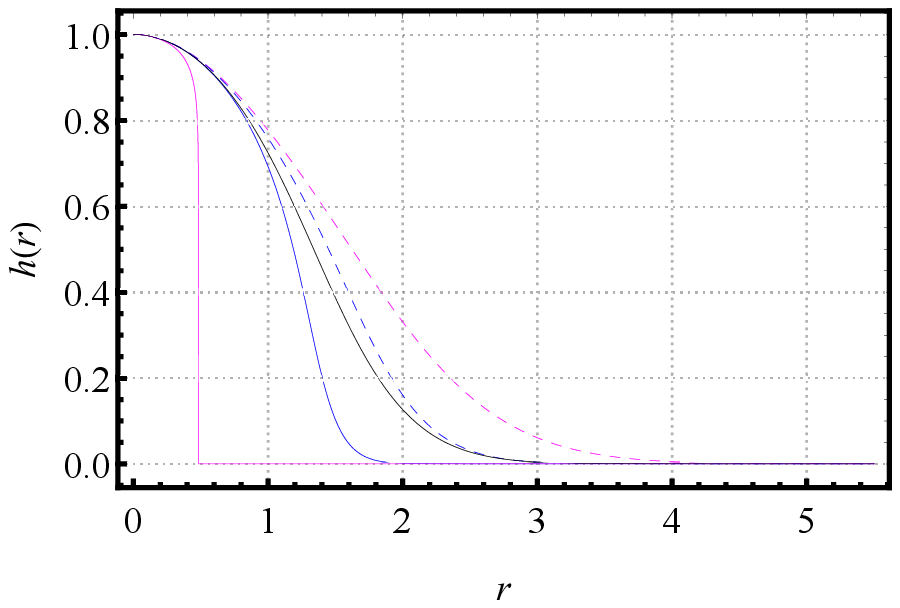}
\caption{Numerical solutions to the Skyrme profile function $h\left(
r\right) $ obtained from {{Eqs.}} (\protect\ref{cx01}) and
(\protect\ref{cx02}) for different values of $\protect\alpha $. Top: $%
\protect\alpha =0.15$ (solid {nave} line), $\protect\alpha =0.25$
(solid orange line), $\protect\alpha =0.50$ (solid red line), $\protect%
\alpha =0.75$ (dashed orange line) and $\protect\alpha =0.85$ (dashed 
{nave} line). Bottom: $\protect\alpha =1.0$ (dashed {magenta}
line), $\protect\alpha =1.5$ (dashed blue line), $\protect\alpha =2.0$
(solid blue line) and $\protect\alpha =4.0$ (solid {magenta} line).
The usual solution appears as a solid black line, for comparison.}
\end{figure}
Thus, the energy density for time-independent fields reads%
\begin{equation}
\varepsilon =-\mathcal{L}\text{,}  \label{energy0}
\end{equation}%
which {can be} written explicitly as {(here, we have already
implemented }$A_{0}=0${)}%
\begin{equation}
\varepsilon =\frac{G}{2g^{2}}B^{2}+\frac{\lambda ^{2}}{2}Q^{2}+V\text{,}
\label{ed00}
\end{equation}%
where the function $Q\equiv \vec{\varphi}\cdot (D_{1}\vec{\varphi}\times
D_{2}\vec{\varphi})$ gives%
\begin{equation}
Q^{2}=\frac{1}{2}(D_{i}\vec{\varphi}\times D_{j}\vec{\varphi})^{2}\text{.}
\end{equation}%
The {finite-energy} requirement }$\varepsilon ({\left\vert \vec{x}%
\right\vert \rightarrow \infty })\rightarrow 0${\ {} establishes
the boundary conditions {which must be} satisfied by the fields of
the model.} 
\begin{figure}[tbp]
\includegraphics[width=8.4cm]{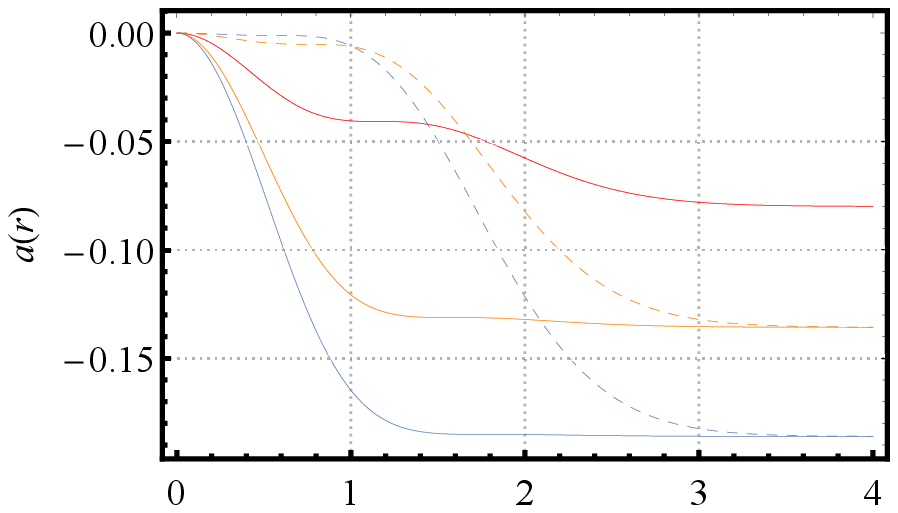}\vspace{0.5cm} %
\includegraphics[width=8.4cm]{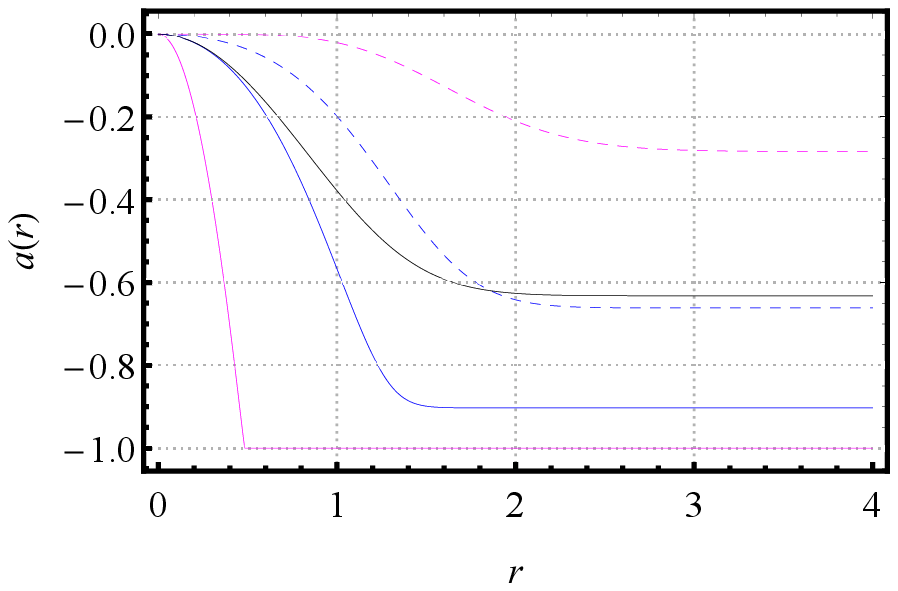}
\caption{Numerical solutions to the gauge profile function $a\left( r\right) 
$ obtained from {Eqs.} (\protect\ref{cx01}) and (\protect
\ref{cx02}) for different values of $\protect\alpha $. Conventions as in the
Fig. 1.}
\end{figure}

{The total energy }$E${\ is defined {as the integral of}
the energy density (\ref{energy0}), so that the implementation of the BPS
formalism allows us to write}%
\begin{eqnarray}
E &=&\int d^{2}\mathbf{x}\left[ \frac{\left( GB\pm \lambda ^{2}g^{2}\mathcal{%
W}\right) ^{2}}{2Gg^{2}}+\frac{\lambda ^{2}}{2}\left( Q\pm Z\right)
^{2}\right.  \notag \\[0.2cm]
&&\left. \mp \lambda ^{2}B\mathcal{W}-\frac{\lambda ^{4}g^{2}}{2G}W^{2}\mp
\lambda ^{2}QZ-\frac{\lambda ^{2}}{2}Z^{2}+V\right] \text{,}\quad \quad
\label{en1}
\end{eqnarray}%
{where we have introduced two auxiliary functions {(to be
determined later below)}, namely $\mathcal{W}\equiv \mathcal{W}(\varphi
_{n}) $ and $Z\equiv Z(\varphi _{n})$.} {Now,} {by using the
expression}%
\begin{equation}
Q=\vec{\varphi}\cdot (\partial _{1}\vec{\varphi}\times \partial _{2}\vec{%
\varphi})+\epsilon _{ij}A_{i}(\widehat{n}\cdot \partial _{j}\vec{\varphi})
\label{Q1}
\end{equation}%
{together with} $B=-\epsilon _{ij}\partial _{i}A_{j}$, we rewrite (%
\ref{en1})\ as%
\begin{eqnarray}
E &=&\int d^{2}\mathbf{x}\left[ \frac{\left( GB\pm \lambda ^{2}g^{2}\mathcal{%
W}\right) ^{2}}{2Gg^{2}}+\frac{\lambda ^{2}}{2}\left( Q\pm Z\right)
^{2}\right.  \notag \\[0.2cm]
&&\mp \lambda ^{2}Z\vec{\varphi}\cdot (\partial _{1}\vec{\varphi}\times
\partial _{2}\vec{\varphi})  \notag \\[0.2cm]
&&\mp \lambda ^{2}\epsilon _{ij}\left[ (\partial _{j}A_{i})\mathcal{W}%
+A_{i}Z(\widehat{n}\cdot \partial _{j}\vec{\varphi})\right]  \notag \\[0.2cm]
&&\left. -\frac{\lambda ^{4}g^{2}}{2G}\mathcal{W}^{2}-\frac{\lambda ^{2}}{2}%
Z^{2}+V\right] \text{.}  \label{enNN}
\end{eqnarray}%
\begin{figure}[tbp]
\includegraphics[width=8.4cm]{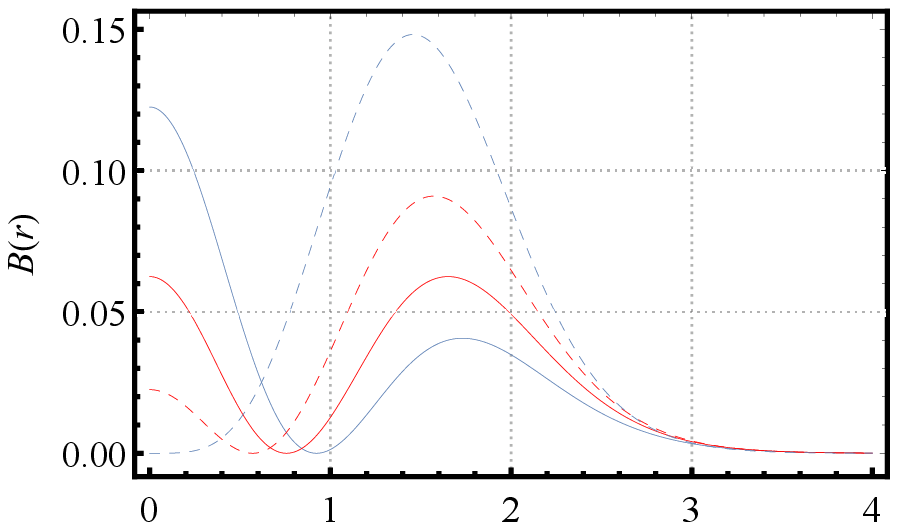}\vspace{0.5cm} \centering%
\includegraphics[width=8.4cm]{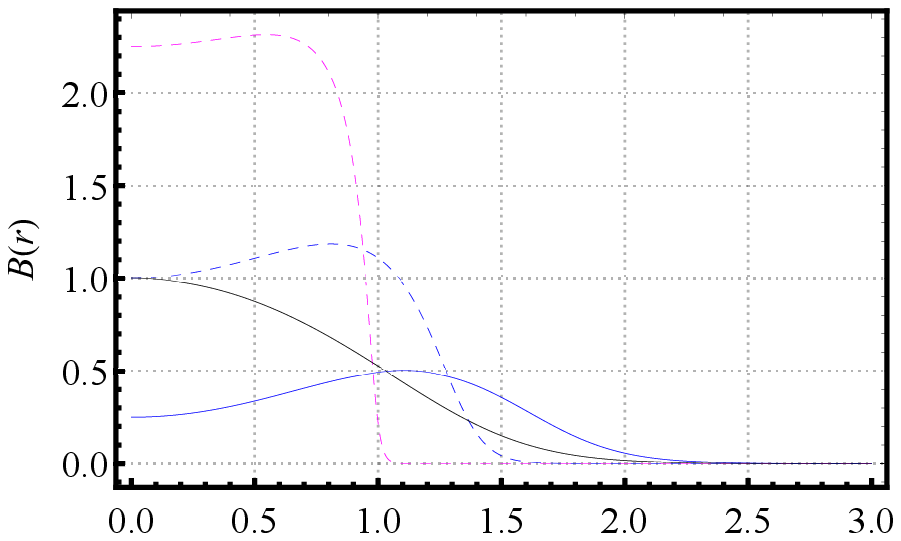}\vspace{0.5cm} %
\includegraphics[width=8.4cm]{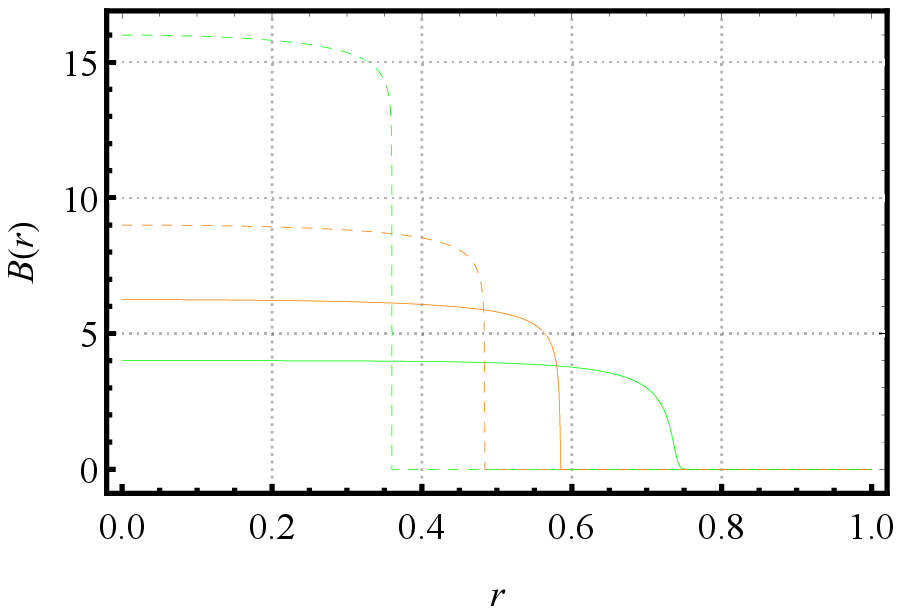}
\caption{Numerical solutions to the BPS magnetic field $B\left( r\right) $
obtained via {Eqs.} (\protect\ref{cx01}) and (\protect\ref%
{cx02}) for different values of $\protect\alpha $. Top: $\protect\alpha %
=0.65 $ (solid {nave} line), $\protect\alpha =0.75$ (solid red line), 
$\protect\alpha =0.85$ (dashed red line) and $\protect\alpha =1$ (dashed 
{nave} line). Middle: $\protect\alpha =1.5$ (solid blue line), $%
\protect\alpha =2$ (dashed blue line) and $\protect\alpha =2.5$ (dashed 
{magenta} line). Bottom: $\protect\alpha =3$ (solid {green}
line), $\protect\alpha =3.5$ (solid orange line), $\protect\alpha =4$
(dashed orange line) and $\protect\alpha =5$ (dashed {green} line).
The usual profile is shown as a solid black line, for comparison.}
\end{figure}
{{At this point,} we transform the third row of {Eq. (%
\ref{enNN})} in a total derivative by choosing%
\begin{equation}
Z=\frac{\partial \mathcal{W}}{\partial \varphi _{n}}\text{,}
\end{equation}%
{such that}%
\begin{equation}
\partial _{j}\mathcal{W}=\frac{\partial \mathcal{W}}{\partial \varphi _{n}}(%
\widehat{n}\cdot \partial _{j}\vec{\varphi})\text{.}
\end{equation}%
{Next,} we set the fourth row of {Eq. (\ref{enNN})} 
{as being} {zero}, {from which} we {get} the BPS
potential $V(\varphi _{n})$ as%
\begin{equation}
V=\frac{\lambda ^{4}g^{2}}{2G}\mathcal{W}^{2}+\frac{\lambda ^{2}}{2}\left( {%
\frac{\partial \mathcal{W}}{\partial \varphi _{n}}}\right) ^{2}\text{.}
\label{Vv}
\end{equation}%
}

{Notably, $\mathcal{W}(\varphi _{n})$ plays the role of a \textit{%
superpotential} function {which} must be {therefore}
constructed (or proposed) such that the self-dual potential $V(\varphi _{n})$
becomes null when $\varphi _{n}\rightarrow 1$ (or $\left\vert \vec{x}%
\right\vert \rightarrow \infty $), in accordance with {the} Eq. (\ref%
{energy0}). Consequently, the following boundary conditions must be
satisfied,}%
\begin{equation}
\lim_{\varphi _{n}\rightarrow 1}{\mathcal{W}}{(\varphi _{n})}=0%
\text{ \ \ and \ \ }\lim_{\varphi _{n}\rightarrow 1}\frac{\partial \mathcal{W%
}}{\partial \varphi _{n}}=0\text{.}  \label{BcVv}
\end{equation}

The total energy {then} becomes%
\begin{eqnarray}
E &=&\int d^{2}\mathbf{x}\left[ \frac{\left( GB\pm \lambda ^{2}g^{2}\mathcal{%
W}\right) ^{2}}{2Gg^{2}}+\frac{\lambda ^{2}}{2}\left( Q\pm \frac{\partial 
\mathcal{W}}{\partial \varphi _{n}}\right) ^{2}\right.  \notag \\[0.2cm]
&&\hspace{-0.75cm}\left. \mp \lambda ^{2}\left( \frac{\partial \mathcal{W}}{%
\partial \varphi _{n}}\right) \vec{\varphi}\cdot (\partial _{1}\vec{\varphi}%
\times \partial _{2}\vec{\varphi})\mp \lambda ^{2}\epsilon _{ij}\partial
_{j}(A_{i}\mathcal{W})\right] \text{.}  \label{enNN0}
\end{eqnarray}

{{In view of} the boundary conditions (\ref{BcVv}), we
observe {that} the contributions {due to} the total
derivatives {{present} in the second row of {Eq. (\ref%
{enNN0})} vanish, from which we can express the total energy as}%
\begin{equation}
E=\bar{E}+E_{bps}\text{,}  \label{en5}
\end{equation}%
where $\bar{E}$ represents {the integral composed} by the
quadratic terms, {i.e.}%
\begin{equation}
\bar{E}=\int d^{2}\mathbf{x}\left[ \frac{\left( GB\pm \lambda ^{2}g^{2}%
\mathcal{W}\right) ^{2}}{2Gg^{2}}+\frac{\lambda ^{2}}{2}\left( Q\pm \frac{%
\partial \mathcal{W}}{\partial \varphi _{n}}\right) ^{2}\right] \text{,}
\label{en4}
\end{equation}%
and $E_{bps}$ defines the energy lower bound, {which reads}%
\begin{equation}
E_{bps}=\mp \lambda ^{2}\int d^{2}\mathbf{x}\left( \frac{\partial \mathcal{W}%
}{\partial \varphi _{n}}\right) \vec{\varphi}\cdot (\partial _{1}\vec{\varphi%
}\times \partial _{2}\vec{\varphi})>0\text{.}  \label{en3}
\end{equation}
}

Here, we point out {that} the term $\vec{\varphi}\cdot (\partial _{1}%
\vec{\varphi}\times \partial _{2}\vec{\varphi})$ is related to the
topological charge (or topological degree, also called winding number) of
the Skyrme field {by means of}%
\begin{equation}
\deg \left[ \vec{\varphi}\right] =-\frac{1}{4\pi }\int d^{2}\mathbf{x~}\vec{%
\varphi}\cdot (\partial _{1}\vec{\varphi}\times \partial _{2}\vec{\varphi})%
\text{,}
\end{equation}%
where $k\in \mathbb{Z\setminus }\left\{ 0\right\} $. 
\begin{figure}[tbp]
\includegraphics[width=8.4cm]{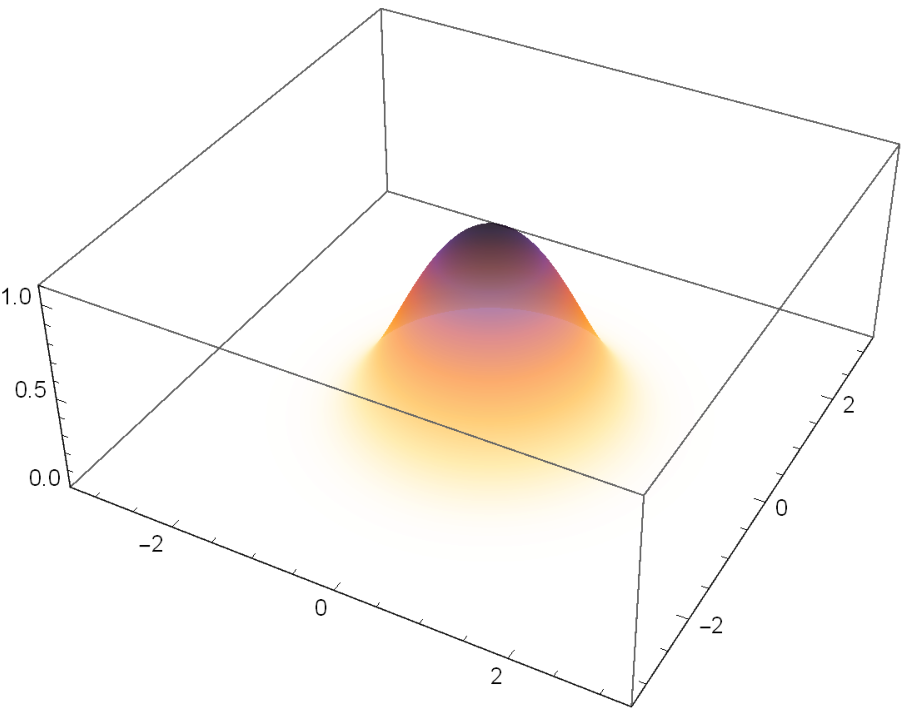}\vspace{0.5cm} %
\includegraphics[width=8.4cm]{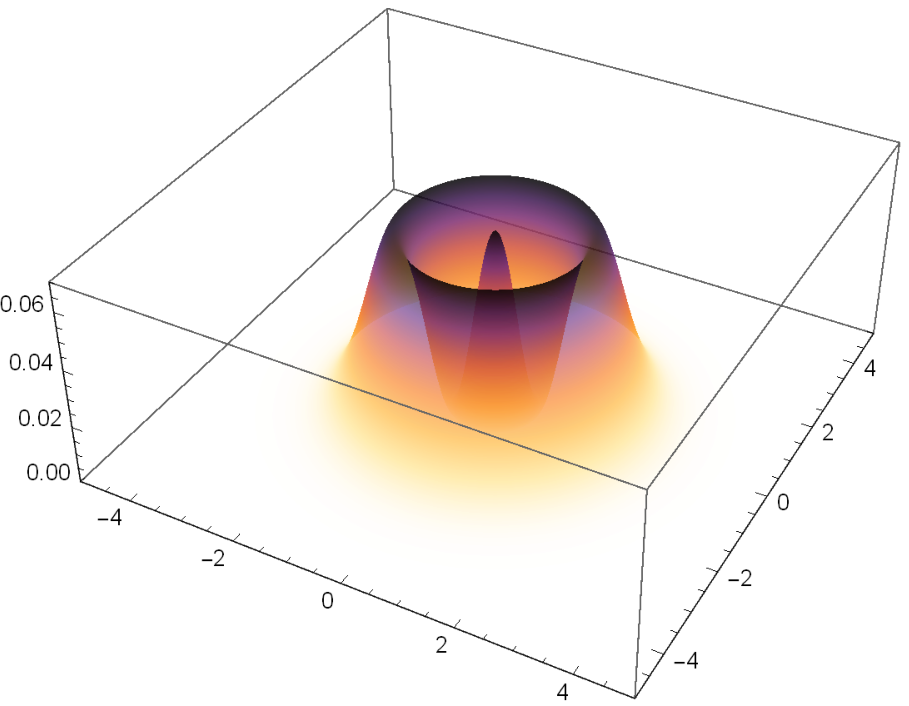}
\caption{Three-dimensional view of the solutions to $B\left( r\right) $.
Top: usual profile. Bottom: solution for $\protect\alpha =0.75$.}
\end{figure}

The total energy (\ref{en5}) {satisfies} the {typical BPS}
inequality%
\begin{equation}
E\geq E_{bps}\text{,}
\end{equation}%
because $\bar{E}\geq 0$. {Then, the energy lower bound (the BPS
one) {is }achieved when the fields {give rise to}
configurations such that $\bar{E}=0$, i.e. when the following set of
first-order differential equations {is} satisfied}:%
\begin{eqnarray}
GB &=&\mp g^{2}\lambda ^{2}\mathcal{W}\text{,}  \label{be10} \\[0.2cm]
Q &=&\mp \frac{\partial \mathcal{W}}{\partial \varphi _{n}}\text{.}
\label{be20}
\end{eqnarray}

Furthermore, from {the combination between} {Eq. (\ref%
{ed00})}, the BPS equations above and the potential (\ref{Vv}), {one
gets that} the BPS energy density {can be} expressed as%
\begin{equation}
\varepsilon _{bps}=\varepsilon _{bps}^{M}+\varepsilon _{bps}^{S}\text{,}
\label{ebps1a}
\end{equation}%
where we have defined%
\begin{eqnarray}
\varepsilon _{bps}^{M} &=&\frac{G}{g^{2}}B^{2}\text{,}  \label{medx0} \\%
[0.2cm]
\varepsilon _{bps}^{S} &=&\lambda ^{2}Q^{2}\text{,}  \label{sedx0}
\end{eqnarray}%
as the \textit{magnetic energy density} and the \textit{Skyrmion energy
density}, respectively. 
\begin{figure}[tbp]
\includegraphics[width=8.4cm]{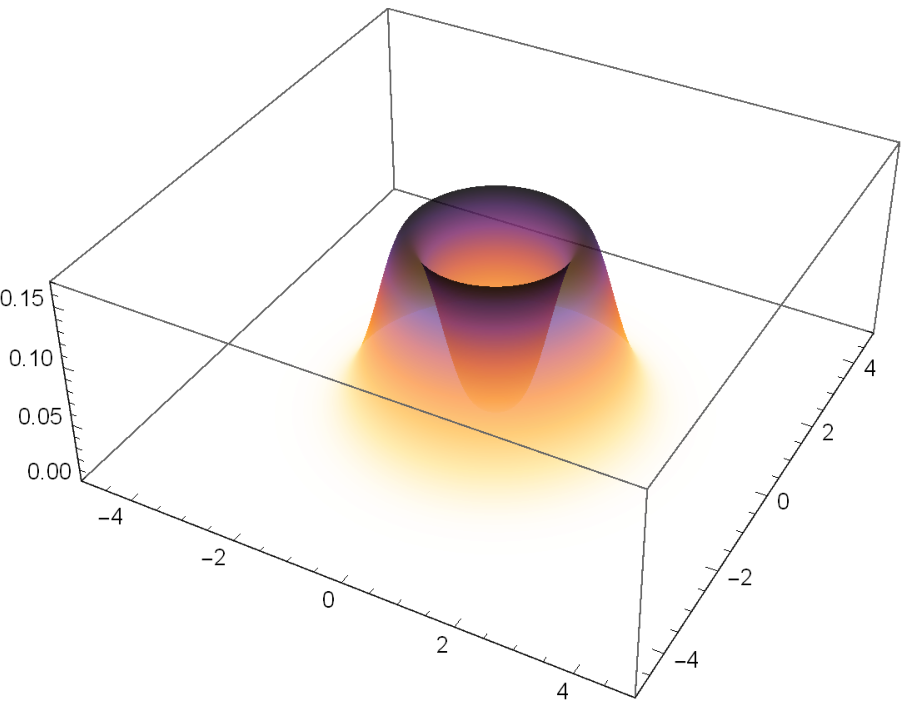}\vspace{0.5cm} %
\includegraphics[width=8.4cm]{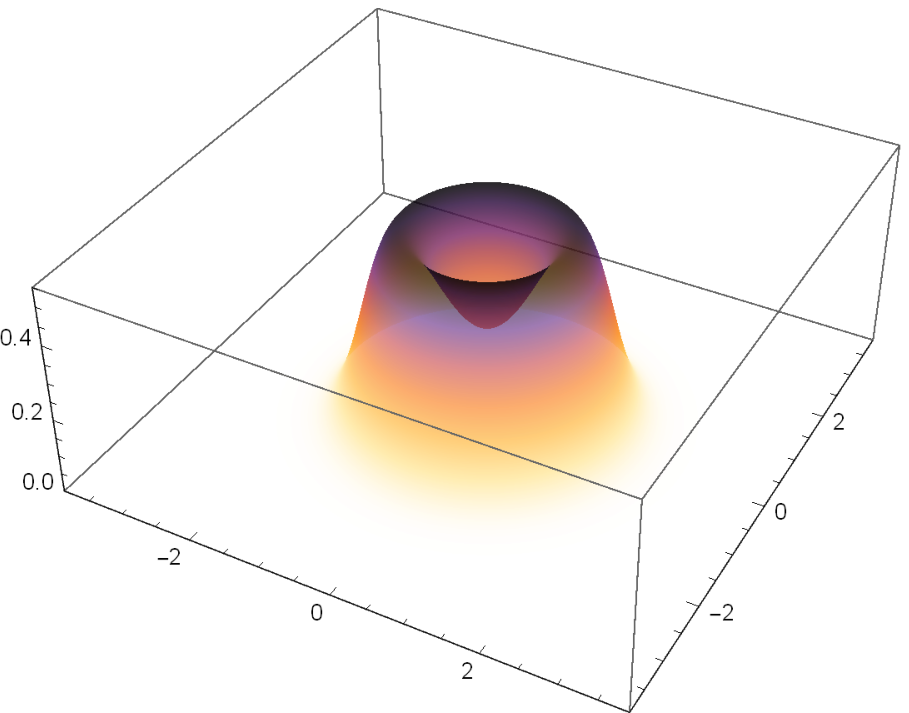}
\caption{Three-dimensional view of the solutions to $B\left( r\right) $. Top
(bottom): solution for $\protect\alpha =1$ ($\protect\alpha =1.5$).}
\end{figure}

{In what follows, we implement the radially symmetric ansatz.}
Without loss of generality, we set $\widehat{n}=\left( 0,0,1\right) $, from
which {}{we get} $\varphi _{n}=\widehat{n}\cdot \vec{%
\varphi}=\varphi _{3}$. In this case, the self-interacting potential $%
V=V\left( \varphi _{3}\right) $ allows {the spontaneous breaking of
the }$SO(3)$ {symmetry} inherent to the Maxwell-Skyrme model (\ref{01}%
), from which configurations with a nontrivial topology are expected to
occur. 
\begin{figure}[tbp]
\includegraphics[width=8.4cm]{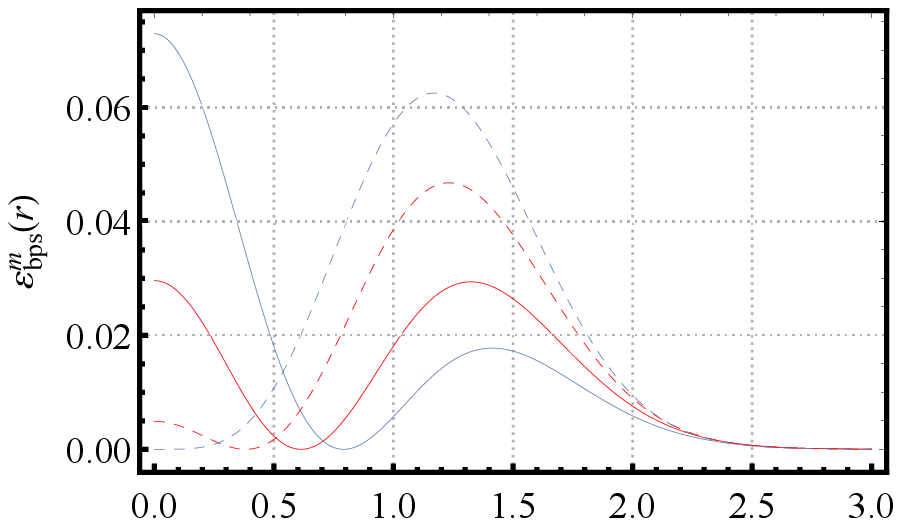}\vspace{0.5cm} %
\includegraphics[width=8.4cm]{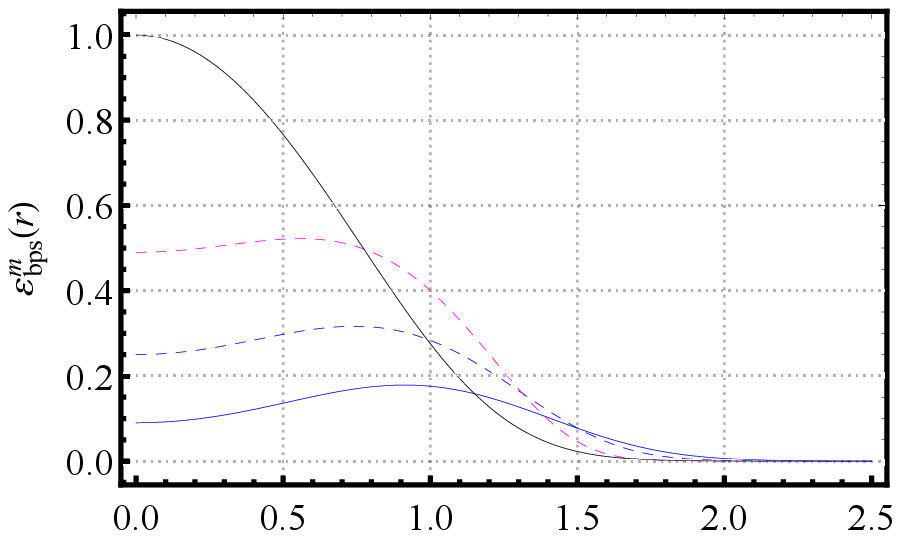}\vspace{0.5cm} %
\includegraphics[width=8.4cm]{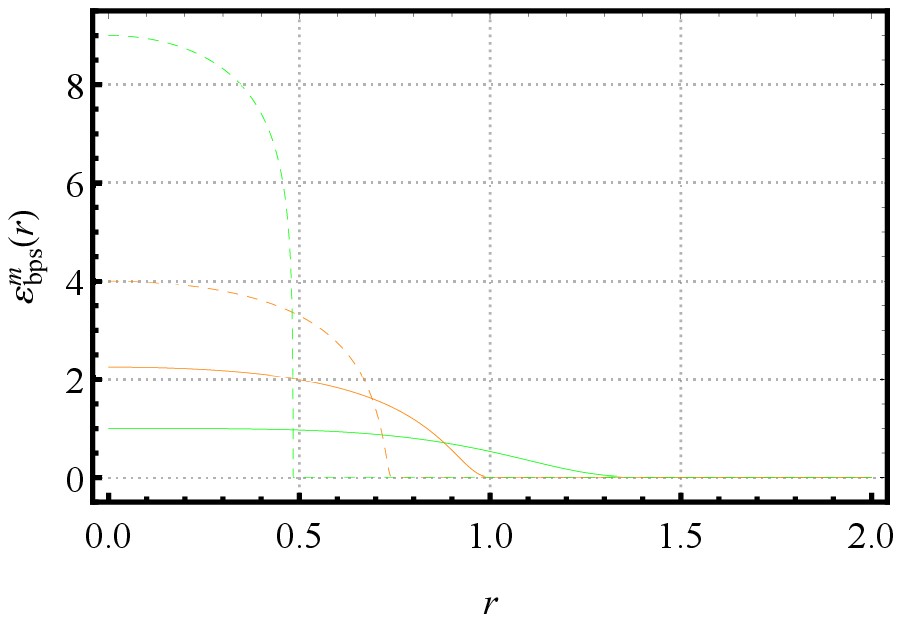}
\caption{Numerical solutions to the BPS magnetic energy density $\protect%
\varepsilon _{bps}^{M}\left( r\right) $ obtained via {Eqs.}
(\protect\ref{cx01}) and (\protect\ref{cx02}) for different values of $%
\protect\alpha $. Top: $\protect\alpha =0.73$ (solid {nave} line), $%
\protect\alpha =2\left( \protect\sqrt{2}-1\right) $ (solid red line), $%
\protect\alpha =0.93$ (dashed red line) and $\protect\alpha =1$ (dashed 
{nave} line). Middle: $\protect\alpha =1.3$ (solid blue line), $%
\protect\alpha =1.5$ (dashed blue line) and $\protect\alpha =1.7$ (dashed 
{magenta} line). Bottom: $\protect\alpha =2$ (solid {green}
line), $\protect\alpha =2.5$ (solid orange line), $\protect\alpha =3$
(dashed orange line) and $\protect\alpha =4$ (dashed {green} line).
The usual profile is shown as a solid black line, for comparison.}
\end{figure}

Moreover, in order to compare our results with the well-established ones, we
study time-independent solutions with radial symmetry through the use of the
standard ansatz, i.e.%
\begin{equation}
A_{i}=-\epsilon _{ij}\widehat{x}_{j}\frac{Na\left( r\right) }{r}\text{,}
\end{equation}%
\begin{equation}
\vec{\varphi}=\left( 
\begin{array}{c}
\sin f\cos \left( N\theta \right) \\ 
\sin f\sin \left( N\theta \right) \\ 
\cos f%
\end{array}%
\right) \text{,}
\end{equation}%
where $r$ and $\theta $ are polar coordinates, $\epsilon _{ij}$ stands for
the antisymmetric symbol (with $\epsilon _{12}=+1$) and $\widehat{x}%
_{i}=(\cos \theta ,\sin \theta )$ represents the unit vector. Also, $N$ is
the winding number of the Skyrme field, while the profile functions $f(r)$
and $a(r)$ are supposed to obey the boundary conditions which are known to
support the existence of regular solutions with finite energy, i.e.%
\begin{eqnarray}
f\left( r=0\right) =\pi \text{ \ } &\text{and}&\text{ \ }f\left(
r\rightarrow \infty \right) \rightarrow 0\text{,}  \label{bc1} \\[0.2cm]
a\left( r=0\right) =0 &\text{ \ and}&\text{ \ }a^{\prime }\left(
r\rightarrow \infty \right) \rightarrow 0\text{,}  \label{bc2}
\end{eqnarray}%
in which prime denotes the derivative with respect to the radial coordinate $%
r$.

The magnetic field in terms of the {radially symmetric} ansatz reads%
\begin{equation}
B(r)=F_{21}=-\frac{N}{r}\frac{da}{dr}\text{.}  \label{mf}
\end{equation}

For convenience, it is useful to implement the field redefinition%
\begin{equation}
h(r)=\frac{1}{2}\left( 1-\cos f\right) \text{,}
\end{equation}%
which satisfy the following boundary conditions,\ 
\begin{equation}
h\left( r=0\right) =1\text{ \ \ and \ \ }h\left( r\rightarrow \infty \right)
\rightarrow 0\text{. }  \label{bc3}
\end{equation}

Consequently, {both} the magnetic permeability and the superpotential
become {functions} of $h$ only,\ i.e. $G\equiv G(h)$\ and $\mathcal{W}%
\equiv \mathcal{W}(h)$, respectively. The boundary conditions satisfied by
the superpotential {can be summarised as}%
\begin{equation}
\lim_{r\rightarrow 0}{\mathcal{W}}{(h)}={\mathcal{W}}_{0}\text{,}\ \
\lim_{r\rightarrow \infty }{\mathcal{W}}{(h)}=0\text{, \ and \ }%
\lim_{r\rightarrow \infty }\frac{\partial \mathcal{W}}{\partial h}=0\text{,}
\label{bbccW}
\end{equation}%
where $\mathcal{W}_{0}=\mathcal{W}(h(r=0))=\mathcal{W}(1)$, whereas the two
last {ones} correspond to {those} {which appear} in {%
Eq. (\ref{BcVv})}.

Further, the magnetic energy density and the Skyrmion energy density {%
can be expressed in terms of the superpotential} {$\mathcal{W}{(h)%
}$} {as} 
\begin{eqnarray}
\varepsilon _{bps}^{M} &=&\lambda ^{4}g^{2}\frac{\mathcal{W}^{2}}{G}\text{,}
\label{medx} \\[0.2cm]
\varepsilon _{bps}^{S} &=&\frac{\lambda ^{2}}{4}\left( \frac{\partial 
\mathcal{W}}{\partial h}\right) ^{2}\text{,}  \label{sedx}
\end{eqnarray}%
respectively.

The BPS energy {given by} {Eq. (\ref{en3})} {can be
calculated explicitly, its value reading}%
\begin{equation}
E_{bps}=\mp 2\pi \lambda ^{2}N{\mathcal{W}}_{0}\text{,}  \label{en3x}
\end{equation}%
{i.e.} a positive-definite quantity (considering ${\mathcal{W}}%
_{0}>0$). Here, the minus (plus) sign corresponds to $N<0$ ($N>0$). 
\begin{figure}[t]
\includegraphics[width=8.4cm]{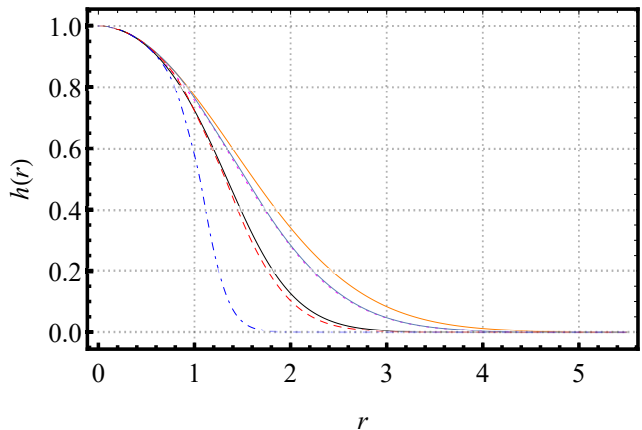}
\caption{Numerical solutions to the Skyrme profile function $h\left(
r\right) $ obtained from the {Eqs.} (\protect\ref{xx3}) and (\protect
\ref{xx3}) for $\protect\alpha =0$ (dotted {magenta} line), $\protect%
\alpha =0.5$ (solid orange line), $\protect\alpha =1$ (solid {nave}
line), $\protect\alpha =1.5$ (dashed red line) and $\protect\alpha =2$
(dotdashed blue line). The usual solution appears as a solid black line, for
comparison.}
\end{figure}

The BPS equations (\ref{be10}) and (\ref{be20}) become%
\begin{equation}
GB=\mp \lambda ^{2}g^{2}\mathcal{W}  \label{be1}
\end{equation}%
and%
\begin{equation}
\frac{\left( 1+a\right) }{r}\frac{dh}{dr}=\pm \frac{1}{4N}\frac{\partial 
\mathcal{W}}{\partial h}\text{,}  \label{be2}
\end{equation}%
respectively. {In summary, these} equations describe a radially
symmetric structure whose total energy is {given by Eq. (\ref%
{en3x})}. {Further,} the BPS gauged skyrmions emerge as the
numerical solutions of the first-order equations (\ref{be1}) and (\ref{be2})
obtained via the boundary conditions (\ref{bc2}) and (\ref{bc3}).

In the next Section, we demonstrate how the first-order framework introduced
above can be used to generate legitimate BPS gauged skyrmions in the
presence of a nontrivial {magnetic permeability}.


\section{BPS Skyrmions in magnetic media\label{sec2}}

The first step to solve the BPS equations is {the specification} 
{of} the superpotential $\mathcal{W}(h)$. {In the present
manuscript, }we use%
\begin{equation}
\mathcal{W}(h)=\frac{h^{2}}{\lambda ^{2}}\text{,}  \label{w1}
\end{equation}%
where $\mathcal{W}_{0}=\lambda ^{-2}$, as desired. {It is clear that}
the superpotential {above} satisfies the conditions given in {%
the} Eq. (\ref{bbccW}). The superpotential (\ref{w1}) {is known to
support} skyrmions which attain their asymptotic values {according to}
a Gaussian decay law, as explained recently in {the Refs.} \cite%
{a2,a3,a4}.

The second step is to choose the function $G(h)$, {i.e.} the magnetic
permeability. We look for gauged {skyrmions} {which} behave
standardly at the boundaries and have a noncanonical {profile} for
intermediate values of the radial coordinate $r$. This type of configuration
was recently investigated by some of us in the context of an extended
Maxwell-$CP(2)$ system, see the Ref. {\cite{ja1}}, for instance. Then, for 
{the present} analysis, we select two different magnetic media, as
shown below.

\subsection{First model}

In order to generate {the aforementioned profiles,} we choose the
magnetic permeability $G(h)$ as 
\begin{equation}
G(h)=\frac{1}{\left( \alpha -h^{2}\right) ^{\beta }}\text{,}  \label{g1}
\end{equation}%
{where $\alpha $ and $\beta $ are positive integer numbers}. In
particular, we intend to clarify how the parameter $\alpha $ changes the
shape of the magnetic field along the radial coordinate, {{%
from which we set} $\beta =2$}. For {the sake of} simplicity, we also
choose the values $g=\lambda =1$ and $N=1$ (i.e. the lower signs in the BPS
equations). 
\begin{figure}[tbp]
\includegraphics[width=8.4cm]{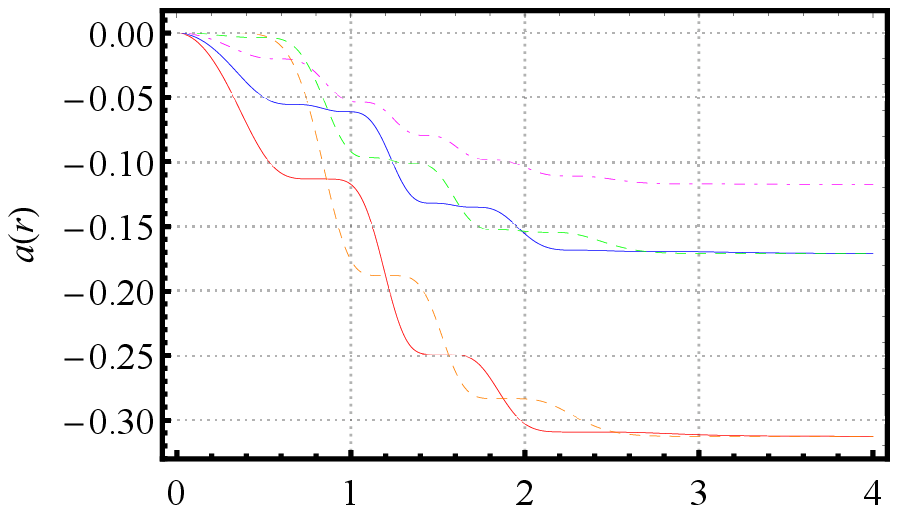}\vspace{0.5cm} %
\includegraphics[width=8.4cm]{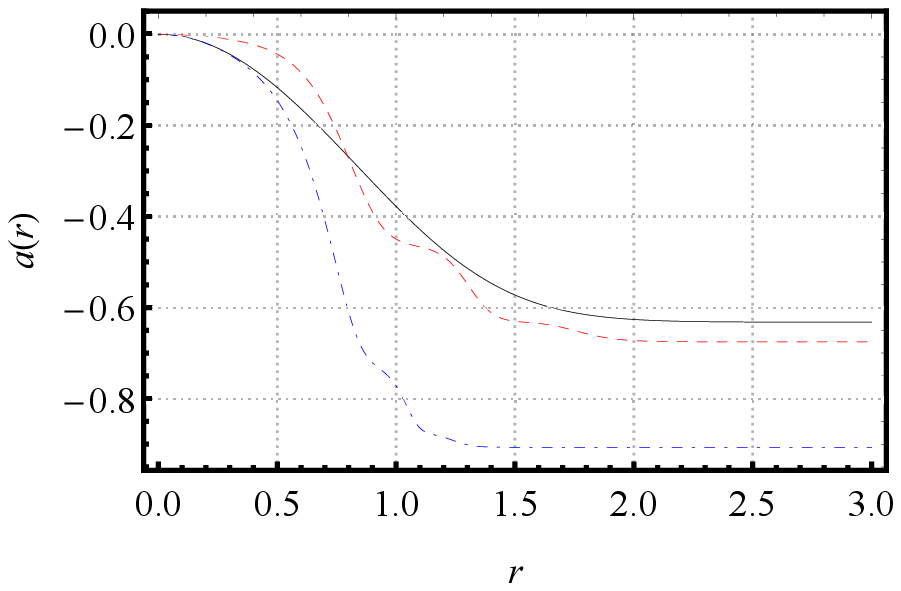}
\caption{Numerical solutions to the gauge profile function $a\left( r\right) 
$ obtained from Eqs. (\protect\ref{xx3}) and (\protect\ref{xx4}) for
different values of $\protect\alpha $. Top: $\protect\alpha =0$ (solid red
line), $\protect\alpha =0.25$\ (solid blue line), $\protect\alpha =0.50$
(dotdashed magenta line), $\protect\alpha =0.75$ (dashed green line) and $%
\protect\alpha =1$ (dashed orange line). Bottom: $\protect\alpha =1,5$
(dashed red line), $\protect\alpha =2$ (dotdashed blue line) and the usual
profile (solid black line).}
\end{figure}

In view of the choices above, the BPS equations (\ref{be1}) and (\ref{be2})
reduce to 
\begin{eqnarray}
\frac{1}{r}\frac{da}{dr} &=&-h^{2}\left( \alpha -h^{2}\right) ^{2}\text{,}
\label{cx01} \\[0.2cm]
\frac{\left( 1+a\right) }{r}\frac{dh}{dr} &=&-\frac{1}{2}h\text{,}
\label{cx02}
\end{eqnarray}%
which we solve numerically via the implementation of a finite-difference
scheme together {with} the boundary conditions (\ref{bc2}) and (\ref%
{bc3}).

In this sense, the figures {1 and 2 show respectively the numerical
profiles to} $h(r)$ {and} $a(r)$ {for different values of} $%
\alpha $. In general, this parameter controls the distance over which those
fields effectively spread, {{therefore} affecting the
intensity of the interaction between these configurations.} Here, it is also
important to note that the solutions to the gauge profile function $a(r)$
with $\alpha <1$ are characterized by the presence of interesting plateaus
which appear for intermediate values of the radial coordinate $r$. For $%
\alpha >3$, both {fields assume a} compact-like {profile} for
increasing $\alpha $. Such a behavior is analogue {to that already
found} in the standard case ($G=1$) for increasing values of the coupling
constant $g$.

As we explain below, these structures give rise to the formation of {%
nonstandard} internal structures which distinguish the behavior of the
corresponding magnetic field, see the analytical considerations in the
sequence.

\subsubsection{The magnetic field}

In the Fig. 3, we show the numerical solutions to the magnetic field $B(r)$,
from which it is possible to see how the shape of this field depends on the
value of $\alpha $ in a dramatic way. We now proceed with an analytical
study of such a dependence.

In order to describe the way the parameter $\alpha $ affects the shape of
the magnetic sector, we write the magnetic field 
\begin{equation}
B(r)=h^{2}\left( \alpha -h^{2}\right) ^{2}\text{,}  \label{be02}
\end{equation}%
whose first derivative provides%
\begin{equation}
\frac{d}{dr}B=2h\left( \alpha -h^{2}\right) \left( \alpha -3h^{2}\right) 
\frac{dh}{dr}\text{.}
\end{equation}

Now, given that the solution to the Skyrme profile function $h(r)$ varies
monotonically from $1$ (at $r=0$) to $0$ (in the limit $r\rightarrow \infty $%
, i.e. the first derivative of the $h(r)$ is always negative), {one
gets that} the condition $B^{\prime }(R)=0$ provides the extreme points $R$
of interest (the ones located away from the origin)%
\begin{eqnarray}
h(R_{1}) &=&h_{1}=\sqrt{\alpha }<1\text{,}  \label{h1} \\[0.2cm]
h(R_{2}) &=&h_{2}=\sqrt{\frac{\alpha }{3}}<1\text{,}  \label{h2}
\end{eqnarray}%
where $0<R_{1}<R_{2}$.

{At these points,} the magnetic field reads%
\begin{eqnarray}
B_{1} &=&B\left( h_{1}\right) =0\text{,}  \label{bb1} \\[0.2cm]
B_{2} &=&B\left( h_{2}\right) =\frac{4}{27}\alpha ^{3}\text{,}  \label{bb2}
\end{eqnarray}%
respectively. The first value, $B_{1}$, becomes a local minimum if $\alpha
<1 $, whereas $B_{2}$ results {in} a local maximum if $\alpha <3$. 
{Moreover,} from {Eq. (\ref{be02})}, the value of the
magnetic field at the origin {is given by}%
\begin{equation}
B_{0}=B\left( r=0\right) =\left( \alpha -1\right) ^{2}\text{.}  \label{bb3}
\end{equation}

In view of the Eqs. (\ref{bb1}), (\ref{bb2}) and (\ref{bb3}) above, we
enumerate {six} different pictures to be considered based on the
values of $\alpha $. In this case, it is important to emphasize that we are
considering all the values of $r$, except the ones located in the {%
asymptotic} region $r\rightarrow \infty $.

\paragraph{The first picture:}

It is defined for $\alpha =0$, from which one gets that the first-order
equations (\ref{cx01}) and (\ref{cx02}) can be rewritten as%
\begin{eqnarray}
\frac{1}{r}\frac{da}{dr} &=&-h^{6}\text{,} \\[0.2cm]
\frac{\left( 1+a\right) }{r}\frac{dh}{dr} &=&-\frac{1}{2}h\text{,}
\end{eqnarray}%
which, after the field redefinition $H(r)=\left[ h(r)\right] ^{3}$, reads%
\begin{eqnarray}
\frac{1}{r}\frac{da}{dr} &=&-H^{2}\text{,}  \label{bx01} \\[0.2cm]
\frac{\left( 1+a\right) }{r}\frac{dH}{dr} &=&-\frac{3}{2}H\text{.}
\label{bx02}
\end{eqnarray}

In this case, despite the redefinition applied on the Skyrme profile
function, we note that the resulting first-order {Eqs.} (\ref%
{bx01}) and (\ref{bx02}) can be obtained directly from the general ones (\ref%
{be1}) and (\ref{be2}) for $G=1$, $g=N=1$, $\sigma =2$ and $\lambda =\sqrt{%
1/3}$. We then conclude that the a priori nontrivial case defined by $%
G(h)=h^{-4}$ stands for a merely redefinition of the usual case (defined by $%
G=1$) with a different value of the coupling constant $\lambda $. As a
consequence, we do not expect significant changes to occur on the shape of
the solutions, especially on that of the magnetic sector. Therefore, in what
follows, we consider only the case with nonvanishing values of $\alpha $.

\paragraph{The second picture:}

It occurs when $0<\alpha <1$. In this context, the solution (\ref{h1}) is
satisfied at some point $r=R_{1}$ (defined via $h\left( r=R_{1}\right)
=h_{1}=\sqrt{\alpha }$). At this point, the magnetic field vanishes (i.e., $%
B\left( r=R_{1}\right) =0$, see {Eq.} (\ref{bb1})), from which it
is reasonable to infer that the magnetic solution describes {a
centered lump surrounded by a ring}: the {lump} is {positioned}
at the origin, its amplitude being given by {Eq.} (\ref{bb3})
itself, while the {radius of the ring} is {located }at some
point $r=R_{2}>R_{1}$ (defined by $h\left( r=R_{2}\right) =h_{2}=\sqrt{%
\alpha /3}$, see {Eq.} (\ref{h2})), {the amplitude of the
ring} standing for $B\left( r=R_{2}\right) =\left( 4/27\right) \alpha ^{3}$,
according to the previous Eq. (\ref{bb2}). 
\begin{figure}[t]
\includegraphics[width=8.4cm]{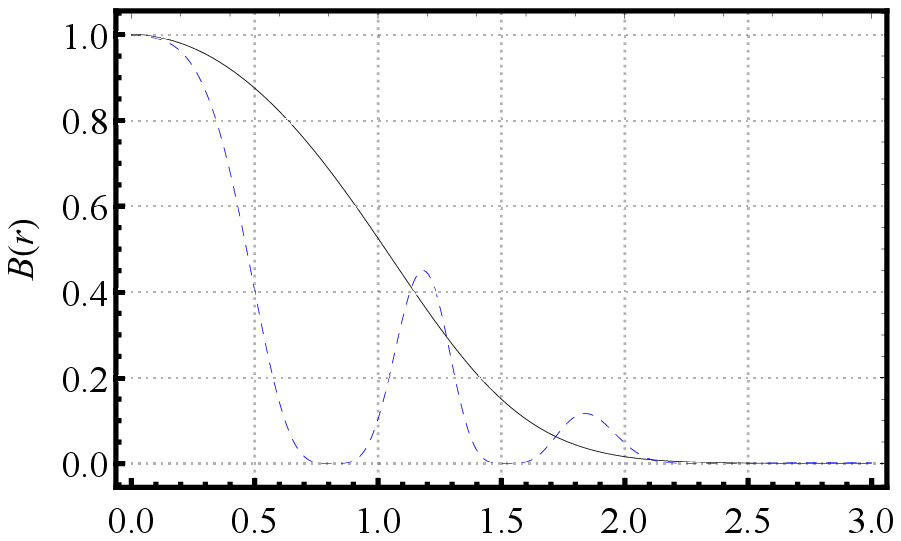}\vspace{0.5cm} %
\includegraphics[width=8.4cm]{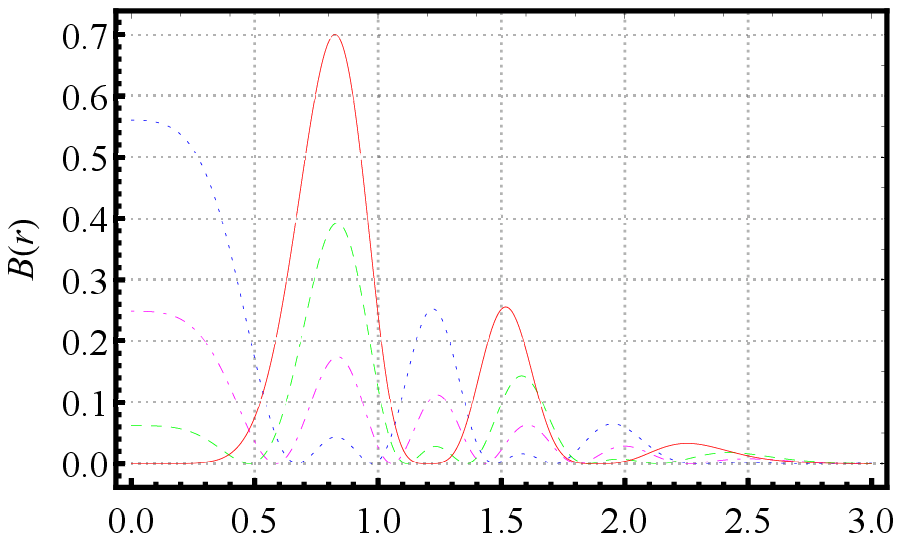}\vspace{0.5cm} %
\includegraphics[width=8.4cm]{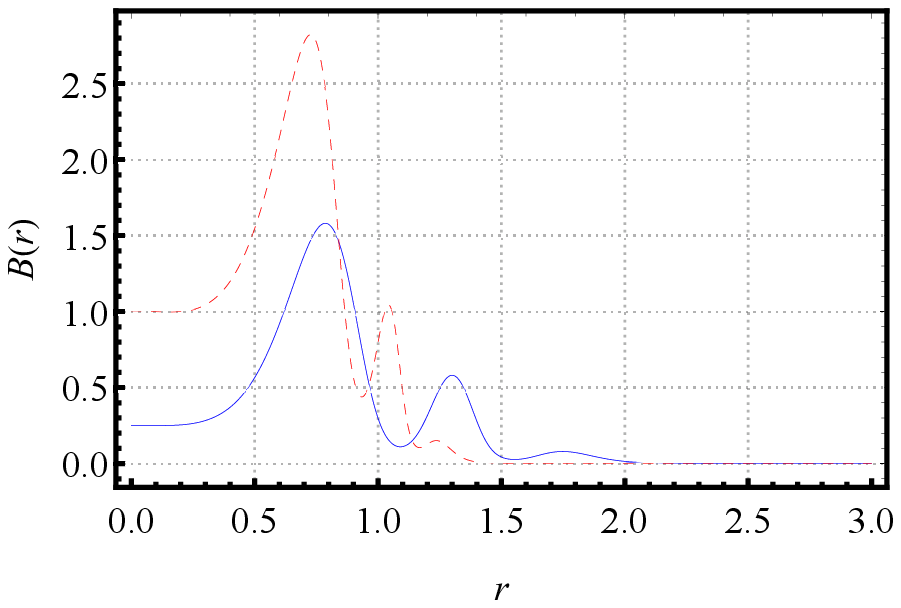}
\caption{Numerical solutions to the BPS magnetic field $B\left( r\right) $
obtained via {Eqs.} (\protect\ref{xx3}) and (\protect\ref{xx4}) for
different values of $\protect\alpha $. Top: $\protect\alpha =0$ (dashed blue
line) and the usual profile (solid black line). Middle: $\protect\alpha %
=0.25 $ (dotted blue line), $\protect\alpha =0.50$ (dotdashed {magenta%
} line), $\protect\alpha =0.75$ (dashed {green} line) and $\protect%
\alpha =1$ (solid red line). Bottom: $\protect\alpha =1.5$ (solid blue line)
and $\protect\alpha =2$ (dashed red line).}
\end{figure}

We highlight how $\alpha $ determines the difference between the amplitudes
of these two {amplitudes}:\ for $0<\alpha <0.75$, {the
magnitude of the centered lump} is taller than {that} {of the
ring} (i.e. $\left( \alpha -1\right) ^{2}>\left( 4/27\right) \alpha ^{3}$).
On the other hand, when $\alpha =0.75$, the two {magnitudes} have the
very same amplitude. Finally, for $0.75<\alpha <1$, the {magnitude of
the ring }is taller than {that of the lump positioned} at $r=0$ (i.e. 
$\left( \alpha -1\right) ^{2}<\left( 4/27\right) \alpha ^{3}$).

The parameter $\alpha $ also controls the values of both $R_{1}$ (i.e. the
point at which $B(r)$ vanishes) and $R_{2}$ (the {radius of the ring}%
): as $\alpha $ increases, the values of $h_{1}=\sqrt{\alpha }$ and $h_{2}=%
\sqrt{\alpha /3}$ also increase and, once $h(r)$ varies monotonically from $%
1 $ to $0$, both $R_{1}$ and $R_{2}$ decrease {(i.e. move toward the
origin)}.

\paragraph{The third picture:}

The case with $\alpha =1$ defines the third picture, for which {%
Eq. (\ref{h1})} holds at the origin only, i.e. $h\left( r=0\right) =h_{1}=1$%
, which agrees with the boundary condition (\ref{bc3}). Therefore, the
magnetic field vanishes at $r=0$, which agrees with the result which comes
from {Eq. (\ref{bb3})} for $\alpha =1$. We then conclude that the
resulting magnetic profile stands for a {single ring} {whose
radius} is located at some point $r=R_{2}$ (defined by $h\left(
r=R_{2}\right) =h_{2}=\sqrt{1/3}$, see the Eq. (\ref{h2})), its magnitude
being equal to $B\left( r=R_{2}\right) =4/27$, see the Eq. (\ref{bb2}).

\paragraph{The fourth picture:}

It is defined for $1<\alpha <3$. In this case, {Eq. (\ref{h1})}
is not satisfied at all, from which one gets that $B(r)$ does not vanish for
intermediate values of $r$ (i.e. {Eq. (\ref{bb1})} does not hold
anymore). The magnetic solution therefore stands for a {volcano}
centered at the origin: the value of the $B(r)$ at $r=0$ is still given by {Eq. (\ref{bb3})}, while {its global maximum} is {%
reached} at $r=R_{2}$ (such that $h\left( r=R_{2}\right) =h_{2}=\sqrt{\alpha
/3}$, for which the magnetic profile {attains} the value given by {%
Eq. (\ref{bb2}))}. The {volcano} is {therefore}
completely characterized by the fact that $B\left( r=R_{2}\right) >B\left(
r=0\right) $. Moreover, $\alpha $ also controls the value of $R_{2}$: when $%
\alpha $ increases, the value of $h(r=R_{2})=\sqrt{\alpha /3}$ also
increases and, once that $h(r)$ varies monotonically from $1$ (at $r=0$) to $%
0$ (in the limit $r\rightarrow \infty $), the value of $R_{2}$ decreases.

\paragraph{The fifth picture:}

We continue our study and define the fifth picture for $\alpha =3$, which
leads us to the conclusion that {Eq. (\ref{h1})} is again not
satisfied at all (i.e. $B(r)$ does not vanish for intermediate $r$). {%
On the other hand,} {Eq. (\ref{h2})} is satisfied only at the
origin (i.e. $h\left( r=0\right) =h_{2}=1$, which agrees with the condition (%
\ref{bc3})). At this point, the Eq. (\ref{bb2}) reveals that the magnetic
field is equal to $B\left( h=h_{2}\right) =4$, which coincides with the
result which arises from {Eq. (\ref{bb3})}. As a consequence, the
magnetic sector describes a compact-like structure centered at $r=0$.

\paragraph{The sixth picture:}

Finally, for $\alpha >3$, both {{Eqs.}} (\ref{h1}) and (%
\ref{h2}) are not satisfied. In such a scenario, one notes that the magnetic
field has no valleys and no additional peaks, i.e. $B(r)$ varies
monotonically from $B\left( r=0\right) =\left( \alpha -1\right) ^{2}$ to $%
B\left( r\rightarrow \infty \right) \rightarrow 0$ (a localized magnetic
flux, as expected). As a result, {as }$\alpha ${\ increases,}
the magnetic field {develops a more and} more compact-like {%
profile} centered at $r=0$. As explained previously, such a behavior is
analogue to {that already found in the} standard case ({with} $%
G=1$) for increasing values of the {electromagnetic} coupling
constant $g$.

\subsubsection{The BPS energy density}

We now focus our attention on the energy density of the resulting BPS
configurations and investigate how this profile is affected by the values of 
$\alpha $. Moreover, in view of our choices (\ref{w1}) and (\ref{g1}) for $%
\mathcal{W}(h)$ and $G(h)$, we get that $\varepsilon _{bps}^{M}$ and $%
\varepsilon _{bps}^{S}$ can be rewritten in the form%
\begin{eqnarray}
\varepsilon _{bps}^{M} &=&h^{4}\left( \alpha -h^{2}\right) ^{2}\text{,}
\label{med} \\[0.2cm]
\varepsilon _{bps}^{S} &=&h^{2}\text{,}
\end{eqnarray}%
where we have already implemented $g=\lambda =1$ and $\beta =2$ (i.e. the
choices inherent to the theoretical scenario which we are effectively
investigating in this manuscript). 
\begin{table}[t]
\caption{Results for $\protect\alpha =0.25$. The exact roots of Eq. (\protect
\ref{eq1}) (first column), the approximate roots of Eq. (\protect\ref{eq2})
(second column) and the values of $B(r)$ calculated at these last roots
(third column). The magnetic field at the origin is given by $B_{0}=0.5625$. The approximate results are given with an accuracy of $10^{-4}$.}%
\begin{ruledtabular}
		\begin{tabular}{clddc}
			{}&Roots of (\protect\ref{eq1})&
			\multicolumn{1}{r}{\textrm{Roots of (\protect
\ref{eq2})}}&{}&{Values of $B(r)$}\\
			\colrule
			& & & & \\ [-0.25cm]
			{}&$8/9$& 0.8350 & &{0.0435} \\		
			{}&$7/9$& 0.6729 & &{0.2524} \\
			{}&$5/9$& 0.5028 & & {0.0157} \\
			{}&$4/9$& 0.3454 & &{0.0648} \\
			{}&$2/9$& 0.1746 & &{0.0018} \\
			{}&$1/9$& 0.0577 & &{0.0008}
		\end{tabular}
	\end{ruledtabular} 
\end{table}

Here, given that the Skyrmion profile $h(r)$ is constrained to satisfy the
boundary conditions (\ref{bc3}) monotonically, it is possible to infer that
the numerical profile for $\varepsilon _{bps}^{S}=h^{2}$ (which does not
depend on $\alpha $ explicitly) stands for a lump centered at $r=0$ for all
values of $\alpha $, from which we {conclude} that there is no
novelty to be discussed concerning this solution.

On the other hand, the solution for the magnetic energy density $\varepsilon
_{bps}^{M}=h^{4}\left( \alpha -h^{2}\right) ^{2}$ depends on $\alpha $
explicitly. It is therefore possible to study the effects caused by
different values of $\alpha $ on the shape of the solution for $\varepsilon
_{bps}^{M}$\ by following the same route as it was done for the magnetic
sector. In this sense, we calculate%
\begin{equation}
\frac{d}{dr}\varepsilon _{bps}^{M}=4h^{3}\left( \alpha -h^{2}\right) \left(
\alpha -2h^{2}\right) \frac{dh}{dr}\text{,}
\end{equation}%
which provides the extreme values {(here, }$0<R_{3}<R_{4}${)}%
\begin{eqnarray}
h(R_{3}) &=&h_{3}=\sqrt{\alpha }<1\text{,}  \label{h3} \\[0.2cm]
h(R_{4}) &=&h_{4}=\sqrt{\frac{\alpha }{2}}<1\text{,}  \label{h4}
\end{eqnarray}%
{via} which one gets the corresponding values of $\varepsilon
_{bps}^{M}$ as%
\begin{eqnarray}
\varepsilon _{bps,3}^{M} &=&\varepsilon _{bps}^{M}\left( h_{3}\right) =0%
\text{,}  \label{med3} \\[0.2cm]
\varepsilon _{bps,4}^{M} &=&\varepsilon _{bps}^{M}\left( h_{4}\right) =\frac{%
\alpha ^{4}}{16}\text{,}  \label{med4}
\end{eqnarray}%
while the value of such density at the origin can be obtained directly from {%
Eq. (\ref{med})}, i.e.%
\begin{equation}
\varepsilon _{bps,0}^{M}=\varepsilon _{bps}^{M}\left( r=0\right) =\left(
\alpha -1\right) ^{2}\text{,}  \label{med0}
\end{equation}%
where we have used $h\left( r=0\right) =1$. 
\begin{table}[b]
\caption{Results for $\protect\alpha =0.50$. Conventions as in the Table I.
The magnetic field at the origin is given by $B_{0}=0.25$. The approximate results are given with an accuracy of $10^{-4}$.}%
\begin{ruledtabular}
		\begin{tabular}{clddc}
			{}&Roots of (\protect\ref{eq1})&
			\multicolumn{1}{r}{\textrm{Roots of (\protect
\ref{eq2})}}&{}&{Values of $B(r)$}\\
			\colrule
			& & & & \\ [-0.25cm]
			{}&$11/12$& 0.8367 & &{0.1743} \\		
			{}&$9/12$& 0.6709 & &{0.1118} \\
			{}&$7/12$& 0.5055 & & {0.0632} \\
			{}&$5/12$& 0.3415 & &{0.0285} \\
			{}&$3/12$& 0.1817 & &{0.0076} \\
			{}&$1/12$& 0.0456 & &{0.0002}
		\end{tabular}
	\end{ruledtabular}
\end{table}

As it was done during the study of the magnetic profile, we again organize
our analysis based on the values of $\alpha $, from which it is possible to
define {six }different pictures to be investigated.

\paragraph{The first picture:}

As we have explained before, it is defined by $\alpha =0$ and can be
interpreted as a redefinition of the usual case with a different value of $%
\lambda $, from which we conclude that this case is not that interesting.
Therefore, in what follows, we consider only those cases with nonvanishing
values of the parameter $\alpha $.

\paragraph{The second picture:}

As before, it comes through $0<\alpha <1$, from which one gets that Eq. (\ref%
{h3}) is satisfied at the point $r=R_{3}$ (i.e. $h\left( r=R_{3}\right)
=h_{3}=\sqrt{\alpha }$). At this position, the magnetic energy density
vanishes (i.e. $\varepsilon _{bps}^{M}\left( r=R_{3}\right) =0$, see Eq. (%
\ref{med3})), from which we infer that the resulting profile stands for a 
{lump (positioned} at $r=0$, its amplitude being equal to $%
\varepsilon _{bps}^{M}\left( r=0\right) =\left( \alpha -1\right) ^{2}$, see
the Eq. (\ref{med0})) {surrounded by a ring (whose radius is defined}
at $r=R_{4}>R_{3}$, with $R_{4}$ defined by $h\left( r=R_{4}\right) =h_{4}=%
\sqrt{\alpha /2}$, see Eq. (\ref{h4})), its amplitude being given by $%
\varepsilon _{bps}^{M}\left( r=R_{4}\right) =\alpha ^{4}/16$, see Eq. (\ref%
{med4}).

We point out that $\alpha $ modulates the difference between the magnitudes
of these two amplitudes, i.e. when $0<\alpha <2\left( \sqrt{2}-1\right) $,
the peak {of the lump} at $r=0$ is taller than {that of the
ring }at $r=R_{4}$ (i.e. $\left( \alpha -1\right) ^{2}>\alpha ^{4}/16$). In
addition, the two peaks have the very same magnitude {for }$\alpha
=2\left( \sqrt{2}-1\right) ${.} However, when $2\left( \sqrt{2}%
-1\right) <\alpha <1$, the peak {of the ring} at $r=R_{4}$ is taller
than {that} located at the origin (i.e. $\left( \alpha -1\right)
^{2}<\alpha ^{4}/16$).

It is clear that $R_{1}=R_{3}$ [note that Eq. (\ref{h1}) is exactly the Eq. (%
\ref{h3})], from which we get that the magnetic field and the magnetic
energy density inevitably vanish at the very same point, see {Eqs.} (%
\ref{be02}) and (\ref{med}), respectively. Moreover, still as a consequence
of $R_{1}=R_{3}$, we conclude that $\alpha $ controls the position of the
valley between the two peaks of the magnetic energy in very same way as it
determines the location of the valley between the two peaks of the magnetic
field itself, see the previous discussion. 
\begin{table}[t]
\caption{Results for $\protect\alpha =0.75$. Conventions as in the Table I.
The magnetic field at the origin is given by $B_{0}=0.0625$. Accuracy of $10^{-4}$.}%
\begin{ruledtabular}
		\begin{tabular}{clddc}
			{}&Roots of (\protect\ref{eq1})&
			\multicolumn{1}{r}{\textrm{Roots of (\protect
\ref{eq2})}}&{}&{Values of $B(r)$}\\
			\colrule
			& & & & \\ [-0.25cm]
			{}&$17/18$& 0.8384 & &{0.3930} \\		
			{}&$13/18$& 0.6688 & &{0.0279} \\
			{}&$11/18$& 0.5083 & & {0.1430} \\
			{}&$7/18$& 0.3375 & &{0.0070} \\
			{}&$5/18$& 0.1884 & &{0.0178} \\
			{}&$1/18$& 0.0314 & &{$3\times 10^{-5}$}
		\end{tabular}
	\end{ruledtabular}
\end{table}

The parameter $\alpha $ controls not only the value of $R_{3}$, but also
that of $R_{4}$ (i.e. {the radius of the ring}): when $\alpha $
increases, the value of $h(r=R_{4})=h_{4}=\sqrt{\alpha /2}$ increases and,
therefore, $R_{4}$ {itself} decreases.

\paragraph{The third picture:}

It is again characterized by $\alpha =1$. In this case, Eq. (\ref{h3}) is
satisfied only at the origin (note that this result agrees with Eq. (\ref%
{bc3})). At this point, the magnetic energy density vanishes, from which we
conclude that the corresponding profile {stands for a single ring
whose global maximum is} located at $r=R_{4}$ (with $R_{4}$ defined by $%
h\left( r=R_{4}\right) =h_{4}=\sqrt{1/2}$, see Eq. (\ref{h4})), {the
maximum itself} being given by $\varepsilon _{bps}^{M}\left( r=R_{4}\right)
=1/16$, see Eq. (\ref{med4}).

\paragraph{The fourth picture:}

It is defined by the values of $\alpha $ within the range $1<\alpha <2$. In
such a scenario, Eq. (\ref{h3}) is not satisfied, from which one gets that
the magnetic energy does not vanish for intermediate values of $r$. As a
consequence, the solution for $\varepsilon _{bps}^{M}\left( r\right) $
results {develops} a {volcano profile} centered at the origin:
the value of that function at $r=0$ is given by the previous Eq. (\ref{med0}%
), with the {global maximum} being located at $r=R_{4}$ (such that $%
h\left( r=R_{4}\right) =h_{4}=\sqrt{\alpha /2}$). At this point, the
magnetic energy density {reaches} the value $\varepsilon
_{bps}^{M}\left( r=R_{4}\right) =\alpha ^{4}/16$, see Eq. (\ref{med4}). The 
{volcano} is then characterized by the fact that $\varepsilon
_{bps}^{M}\left( r=R_{4}\right) >\varepsilon _{bps}^{M}\left( r=0\right) $.

The parameter $\alpha $ also controls the value of $R_{4}$ itself, i.e. when 
$\alpha $ increases, the value of $h(r=R_{4})=\sqrt{\alpha /2}$ increases,
while $R_{4}$ {moves toward} $r=0$.

\paragraph{The fifth picture:}

It is defined by $\alpha =2$. Again, Eq. (\ref{h3}) {does} not hold
anymore (i.e. $\varepsilon _{bps}^{M}\left( r\right) $ does not vanish for
intermediate $r$). Moreover, Eq. (\ref{h4}) is satisfied only at $r=0$,
where the magnetic energy is equal to the unity (this result coincides with
the one which comes from Eq. (\ref{med0}) for $\alpha =2$). We then conclude
that the magnetic energy profile stands for a {lump with a compact
profile} centered at the origin.

\paragraph{The sixth picture:}

We end our analysis by considering the picture defined for $\alpha >2$, for
which none of {Eqs.} (\ref{h3}) and (\ref{h4}) is satisfied. This
fact reveals that the solution for $\varepsilon _{bps}^{M}\left( r\right) $
varies monotonically (i.e. with no valleys and no additional peaks) from $%
\varepsilon _{bps}^{M}\left( r=0\right) =\left( \alpha -1\right) ^{2}$ to $%
\varepsilon _{bps}^{M}\left( r\rightarrow \infty \right) \rightarrow 0$. As
a {consequence,} the resulting configuration {(i.e. a lump)} 
{develops a more and} more compact-like {profile} centered at $%
r=0$ whenever $\alpha $ increases. A similar situation happens for
increasing values of $g$ in the standard case ($G=1$).

\subsection{Second model}

The previous idea about gauged skyrmions with internal structures can be
generalized to include a BPS magnetic configuration with multiple zeros. In
order to illustrate this possibility, we choose the {magnetic
permeability} as%
\begin{equation}
G(h) =\frac{1}{\left[ \alpha -\cos ^{2}\left( n\pi h\right) \right] ^{\beta }%
}\text{,}  \label{g2}
\end{equation}%
where both $\alpha $\ and $\beta $ are again positive integer numbers.

Eq. (\ref{g2}) reveals that, given a particular value of $\alpha $ such that 
$0\leq \alpha \leq 1$, the integer parameter $n$ counts the number of
singularities which characterize the {function }$G(h)$ and therefore
it is expected to define also the number of zeros in the resulting magnetic
solution. {It is important to highlight that a magnetic BPS\ soliton
with multiple zeros was recently found by Bazeia et al. in the context of an
enlarged Maxwell-Higgs theory {\cite{bazeia1}} and that such a profile can
be used as an attempt to explain the behaviour of radially symmetric
configurations at the nanometric scale.}

In view of Eq. (\ref{g2}), we focus our attention on the effects which the
values of $\alpha $ promote on the shape of the resulting solutions, so we
set $\beta =2$ and $n=3$. Furthermore, as previously, we fix $g=\lambda =1 $
and $N=1$, {from which we get that }the first-order equations (\ref%
{be1}) and (\ref{be2}) assume the form%
\begin{eqnarray}
&\displaystyle\frac{1}{r}\frac{da}{dr}=-h^{2}\left[ \alpha -\cos ^{2}\left(
3\pi h\right) \right] ^{2}\text{,}&  \label{xx3} \\[0.2cm]
&\displaystyle\frac{\left( 1+a\right) }{r}\frac{dh}{dr}=-\frac{1}{2}h\text{,}%
&  \label{xx4}
\end{eqnarray}%
{which} must be solved by means of the usual finite-difference scheme
according to the conditions (\ref{bc2}) and (\ref{bc3}).

The figures 7, 8, 9 and 12\ show the numerical results obtained for
different values of $\alpha $. As before, in all these figures, the standard
profile (i.e. the one for $G=1$) is depicted as a solid black line, for
comparison.

In particular, the figures 7 and 8 present the solutions to $h(r)$\ and $a(r)
$, respectively, from which we see that, as in the case studied previously,
different values of $\alpha $ in general change the length over which the
cores of $h(r)$ and $a(r)$ spread. It is also interesting to note the
formation of multiple plateaus in the solutions for the gauge profile
function $a(r)$: as the reader can infer (based on the case investigated
above), these plateaus indicate the existence of a magnetic field
characterized by an internal structure with multiple zeros (which we again
explain theoretically based on the values of $\alpha $\ below).

\subsubsection{The magnetic field}

The Figure 9 {depicts} the profiles {to} the magnetic field $%
B(r)$, from which the reader can verify how the shape of this field changes
dramatically as $\alpha $ {itself varies}. As before, we now proceed
with an analytical investigation on such a relationship, the starting-point
being the radially symmetric expression for the {effective}
first-order BPS magnetic field, i.e.%
\begin{equation}
B=h^{2}\left[ \alpha -\cos ^{2}\left( 3\pi h\right) \right] ^{2}\text{,}
\label{m0}
\end{equation}%
which was obtained through the combination between the equations (\ref{be1}%
), (\ref{w1}) and (\ref{g2}) and the conventions which we have adopted in
this manuscript {regarding} the parameters of the model. 
\begin{figure}[tbp]
\includegraphics[width=8.4cm]{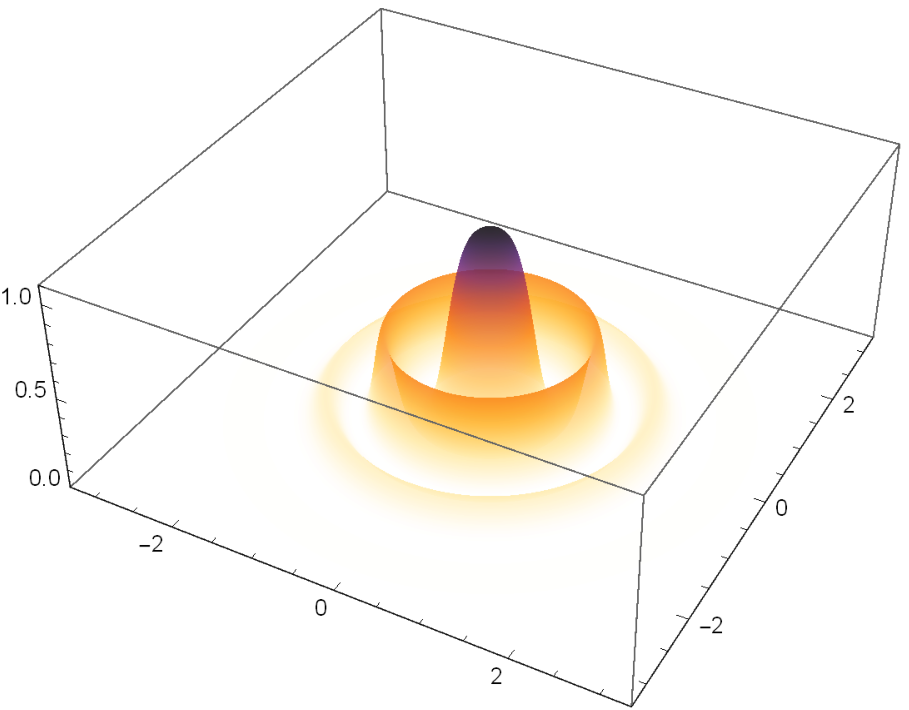}\vspace{0.5cm} %
\includegraphics[width=8.4cm]{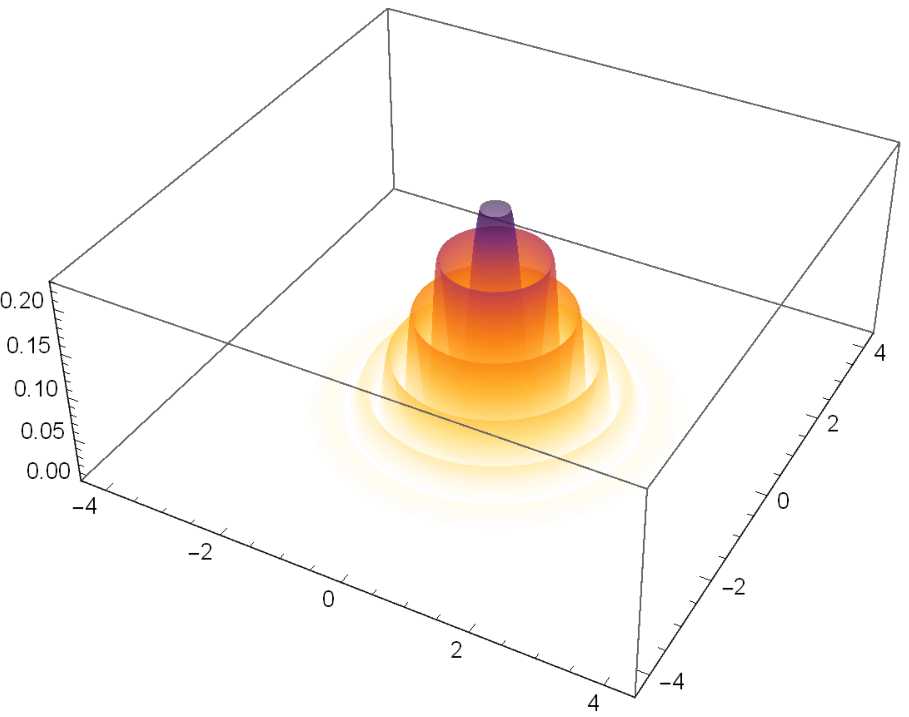}
\caption{Three-dimensional view of the solutions to $B\left( r\right) $. Top
(bottom): solution for $\protect\alpha =0$ ($\protect\alpha =0.5$).}
\end{figure}

{From Eq. (\ref{m0}), we obtain} 
\begin{eqnarray}
\frac{d}{dr}B &=&2h\left[ \alpha -\cos ^{2}\left( 3\pi h\right) \right] 
\notag \\[0.08in]
&&\hspace{-1cm}\times \left[ \frac{{}}{{}}[6\pi h\sin \left( 3\pi h\right)
-\cos \left( 3\pi h\right) ]\cos \left( 3\pi h\right) +\alpha \right] \frac{%
dh}{dr}\text{,}  \label{dB}
\end{eqnarray}%
from which one gets that $B^{\prime }(r)=0$ gives rise to ($h(r)$ is
constrained to vary from 1 to 0 monotonically)%
\begin{eqnarray}
&\cos ^{2}\left( 3\pi h\right) =\alpha \text{,}&  \label{eq1} \\[0.2cm]
&\left[ 6\pi h\sin \left( 3\pi h\right) -\cos \left( 3\pi h\right) \right]
\cos \left( 3\pi h\right) =-\alpha \text{,}&  \label{eq2}
\end{eqnarray}%
where, as before, we are considering all the values of $r$, except those
ones in $r\rightarrow \infty $.

In addition, the value of $B(r)$ at the origin can be obtained from Eq. (\ref%
{m0}), i.e. 
\begin{equation}
B_{0}=\left( \alpha -1\right) ^{2}\text{,}  \label{b0}
\end{equation}%
where we have again used the definition $B_{0}=B(r=0)$.

The reader is expected to infer that from this point on the analysis follows
the same route already stated during the investigation of the previous case,
i.e. for a particular value of $\alpha $, Eqs. (\ref{eq1}) and (\ref{eq2})
must be solved for the values of $h_{i}$. In the sequence, the magnetic
field (\ref{m0}) for that particular $\alpha $ must be evaluated at those
different $h_{i}$'s and its resulting values must be classified as local
maxima or minima (including zero) of the corresponding solution. Based on
this classification, it must be possible to label the magnetic solution as a
lump or as a ring and also to describe theoretically how its main dimensions
(such as amplitude and radius) depend on the value of $\alpha $. 
\begin{figure}[tbp]
\includegraphics[width=8.4cm]{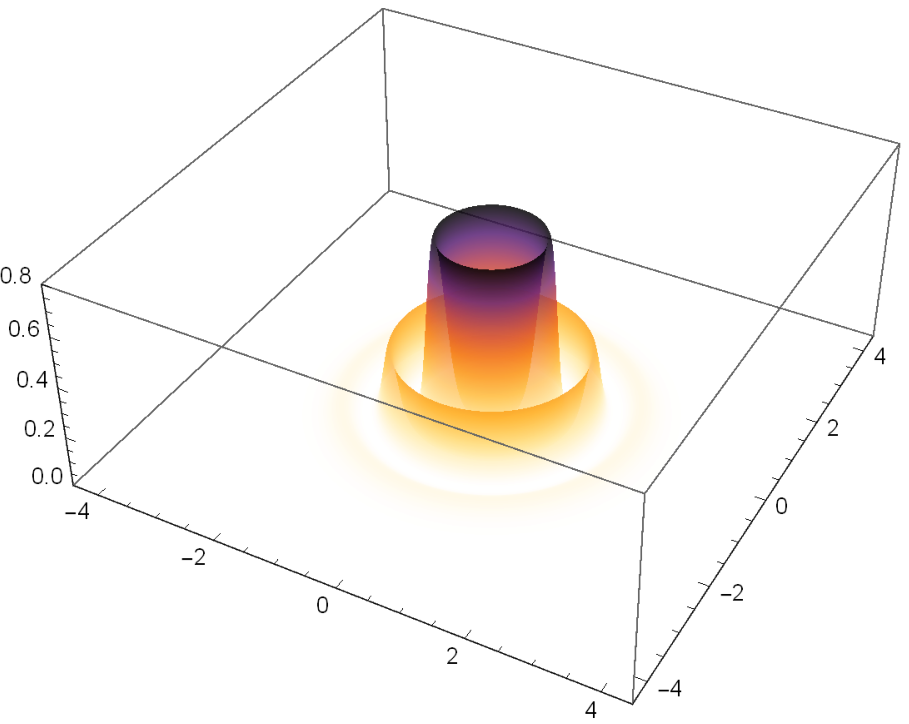}\vspace{0.5cm} %
\includegraphics[width=8.4cm]{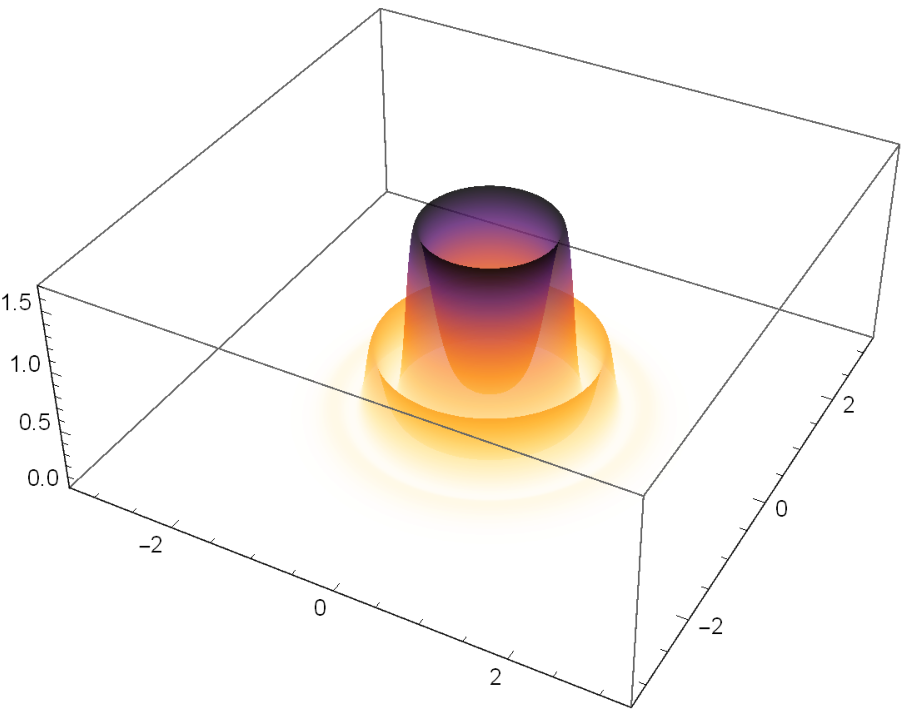}
\caption{Three-dimensional view of the solutions to $B\left( r\right) $. Top
(bottom): solution for $\protect\alpha =1$ ($\protect\alpha =1.5$).}
\end{figure}

Once that we have explained how to proceed, below we use such a prescription
to study {some} ranges of values for $\alpha $ in order to illustrate
our results.

\paragraph{The case $\protect\alpha=0$:}

In this case, Eq. (\ref{eq1}) can be reduced to%
\begin{equation}
\cos \left( 3\pi h\right) =0\text{,}  \label{eq1a}
\end{equation}%
whose exact roots can be easily verified to be%
\begin{eqnarray}
h(\mathcal{R}_{1}) &=&h_{1}=\frac{5}{6}\text{,}  \label{H3} \\[0.2cm]
h(\mathcal{R}_{3}) &=&h_{3}=\frac{3}{6}\text{,}  \label{H2} \\[0.2cm]
h(\mathcal{R}_{5}) &=&h_{5}=\frac{1}{6}\text{,}  \label{H1}
\end{eqnarray}%
{from which one calculates }$B_{1}=B_{2}=B_{3}=0${\ (with }$%
B_{i}=B(h=h_{i})${), i.e.} the points {above} represent $3$
(three) different zeros of the magnetic field.

On the other hand, for $\alpha =0$, Eq. (\ref{eq2}) can be {written as%
}%
\begin{equation}
\left[ 6\pi h\sin \left( 3\pi h\right) -\cos \left( 3\pi h\right) \right]
\cos \left( 3\pi h\right) =0\text{,}  \label{eq20}
\end{equation}%
which promptly gives%
\begin{eqnarray}
&\cos \left( 3\pi h\right) =0\text{,}&  \label{eq2a} \\[0.2cm]
&6\pi h\sin \left( 3\pi h\right) -\cos \left( 3\pi h\right) =0\text{.}&
\label{eq2b}
\end{eqnarray}

The equation (\ref{eq2a}) is the same Eq. (\ref{eq1a}), from which we get
that Eq. (\ref{eq20}) is satisfied by the values of\ $h$ given by {%
Eqs.} (\ref{H1}), (\ref{H2}) and (\ref{H3}) for which $B(r)$ vanishes, see
the discussion therein.

In addition, Eq. (\ref{eq2b}) stands for a transcendental one whose roots
must be therefore evaluated numerically. In this context, there are three
different solutions, i.e.%
\begin{eqnarray}
h_{2}(r &=&\mathcal{R}_{2})\approx 0.6750\text{ \ (}B_{2}\approx 0.4500\text{%
),}  \label{bx1} \\[0.2cm]
h_{4}(r &=&\mathcal{R}_{4})\approx 0.3493\text{ \ (}B_{4}\approx 0.1166\text{%
),}  \label{bx2} \\[0.2cm]
h_{6}(r &=&\mathcal{R}_{6})\approx 0.0693\text{ \ (}B_{6}\approx 0.0019\text{%
),}  \label{bx3}
\end{eqnarray}%
via which {we get} that the {corresponding} values of $B(r)$
stand for the local maxima when $\alpha =0$.

Moreover, the value of the magnetic field at the origin is given by%
\begin{equation}
B_{0}=1\text{,}  \label{p0}
\end{equation}%
according {to} the previous Eq. (\ref{b0}).

The results calculated above reveal that the corresponding magnetic profile
presents a global maximum at $r=0$ and a much more sophisticated internal
structure {which} interpolates between three different zeros [whose
locations are defined by {Eqs.} (\ref{H3}), (\ref{H2}) and (\ref{H1}%
)] and the three different local maxima (with the respective localisations)
given by {Eqs.} (\ref{bx1}), (\ref{bx2}) and (\ref{bx3}), before
finally vanishing in the asymptotic limit{, the resulting profile
therefore standing for a centered lump surrounded by three concentric rings.}

\paragraph{For $0<\protect\alpha <1$:}

In this case, Eq. (\ref{eq1}) predicts the existence of $6$ (six) different
roots whose exact values can be calculated directly from%
\begin{equation}
h_{i,\pm }(r=\mathcal{R}_{i,\pm })=\frac{\arccos \left( \pm \sqrt{\alpha }%
\right) }{3\pi }+\frac{\left( i-1\right) }{3}\text{,}  \label{hi}
\end{equation}%
where $i=1$, $2$ and $3$.

At these six points, the magnetic field vanishes identically, from which one
gets that, for a particular value of $\alpha $ such that $0<\alpha <1$, $%
B(r) $ presents $6$ (six) different zeros.

In addition, Eq. (\ref{eq2}) again does not support an exact solution and
therefore must be solved numerically for different values of $\alpha $
within the range $0<\alpha <1$. The approximate roots obtained this way
eventually define the peaks inherent to the internal structure which
characterizes the magnetic solution. {As before, the value of $%
B(r)\ $at $r=0$ is still given by Eq. (\ref{b0})}.

We\ illustrate\ our analysis by plotting the solutions for the values $%
\alpha =0.25$, $\alpha =0.50$ and $\alpha =0.75$, from which we summarize
the corresponding results in the tables I, II and III below, respectively.
In these tables, the values of $h_{i,\pm }$ (which appear in the first
column) stand for the exact roots of Eq. (\ref{eq1}) (i.e. the points at
which the magnetic field vanishes), while the values of $h_{j}$ (displayed
on the second column) are the approximate roots of the transcendental Eq. (%
\ref{eq2}). Finally, the third column shows the values of $B(r)$ calculated
at these various $h_{j}$, i.e. $B_{j}=B(h=h_{j})$ (the peaks of the magnetic
profile). {In view of these results, we conclude that the resulting
profile represents a lump centered at the origin now surrounded by six
concentric rings.}

In general, the magnetic field for $0<\alpha <1$ possesses a nonvanishing
value $B_{0}$ {(i.e. the magnitude of the lump)} at the origin which
eventually stand for the global maximum of the corresponding solution, see
the results for both $\alpha =0.25$ and $\alpha =0.50$, for instance.
Moreover, for intermediate values of $r$, the magnetic profile develops an
intricate structure which now interpolates between six zeros and six peaks 
{(which define each one of the concentric rings)} before vanishing in
the asymptotic limit.

In particular, based on the results in the tables I, II and III, it is
possible to conclude that the positions of the first, third and fifth
(second, fourth and sixth) zeros of the magnetic field move towards
(outwards) the origin as $\alpha $ increases, with the positions of the 
{rings} inherent to its internal structure behaving in the very same
way. In addition, the amplitudes of the first, third and fifth (second,
fourth and sixth) peaks get higher (lower) as $\alpha $ increases.

\paragraph{The case $\protect\alpha=1$:}

The interested reader can follow this route in order to describe the shape
of the magnetic field for different values of the parameter $\alpha $. For
instance, when $\alpha =1$, Eq. (\ref{eq1}) predicts the existence of four
exact roots, but only two of them are located at intermediate values of $r$
(while the other two roots are positioned at the boundaries). We therefore
conclude that the corresponding magnetic solution presents two different
zeros (beyond the ones located at $r=0$ and in the asymptotics). In
addition, the transcendental Eq. (\ref{eq2}) leads to seven roots, but four
of them are exactly the ones previewed by Eq. (\ref{eq1}) itself, while the
remaining three roots define the peaks {(rings)} which form the
resulting internal structure. In summary, the magnetic solution for $\alpha
=1$\ has a dramatically new profile which vanishes at $r=0$\ (i.e. $B_{0}=0$%
, see Eq. (\ref{b0})) and presents an internal structure formed by three
rings \textit{without} a lump inside them.

\paragraph{The case $\protect\alpha >1$:}

{On the other hand,} when $\alpha =1.5$ and $\alpha =2$, Eq. (\ref%
{eq1}) does not admit any solution, {while} Eq. (\ref{eq2}) itself
provides five (with a sixth one located at $r=0$) approximate roots which
stand for two local minima (not equal to zero) and three local maxima of $%
B(r)$, from which we infer the existence of an internal structure with a new
profile {whose interpolation does not include} different zeros 
{and therefore describes a configuration with three concentric
volcanos.}

As $\alpha $ increases, the number of roots (i.e. the number of minimums and
maximums over which the internal structure interpolates) provided by the
transcendental Eq. (\ref{eq2}) decreases. The interested reader can verify
that, when $\alpha $ is sufficiently large, that Equation does not support
any root, from which the corresponding magnetic field stands for a {%
single} lump centered at the origin which vanishes monotonically in the
asymptotic limit.

\subsubsection{The BPS energy density}

We end our manuscript by clarifying how the magnetic energy distribution
changes with $\alpha $ when the {magnetic permeability} is given by
Eq. (\ref{g2}). In this case, {the magnetic energy density} (\ref%
{medx}) can be written as%
\begin{equation}
\varepsilon _{bps}^{M}=h^{4}\left[ \alpha -\cos ^{2}\left( 3\pi h\right) %
\right] ^{2}\text{,}  \label{med1}
\end{equation}%
where we have implemented our choices to both $\mathcal{W}(h)$ and $G(h)$,
together with $g=\lambda =1$, $\sigma =\beta =2$ and $n=3$.

As in the previous case, the Skyrme energy density (\ref{sedx}) is again
given by $\varepsilon _{bps}^{S}=h^{2}$ and therefore stands for a lump
centered at $r=0$ for all values of $\alpha $ (i.e. no significant
variations to be considered). 
\begin{figure}[t]
\includegraphics[width=8.4cm]{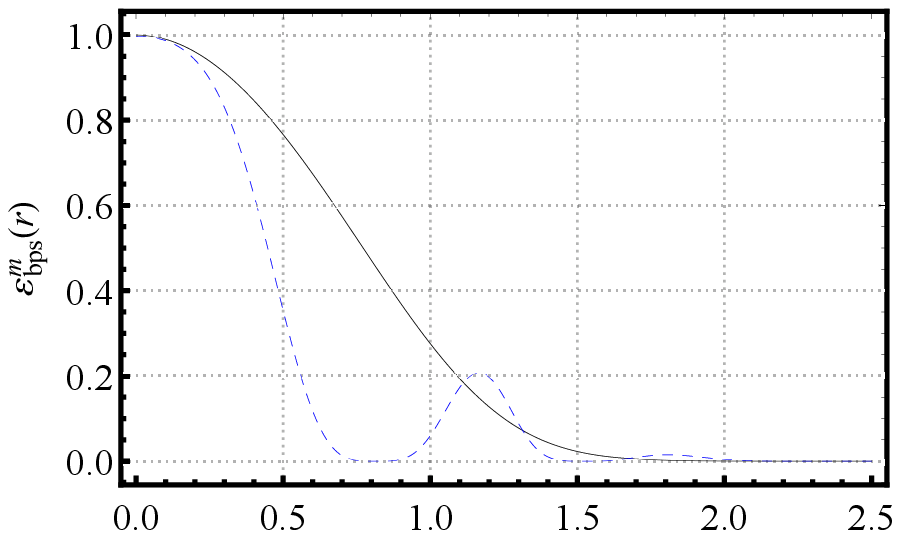}\vspace{0.5cm} %
\includegraphics[width=8.4cm]{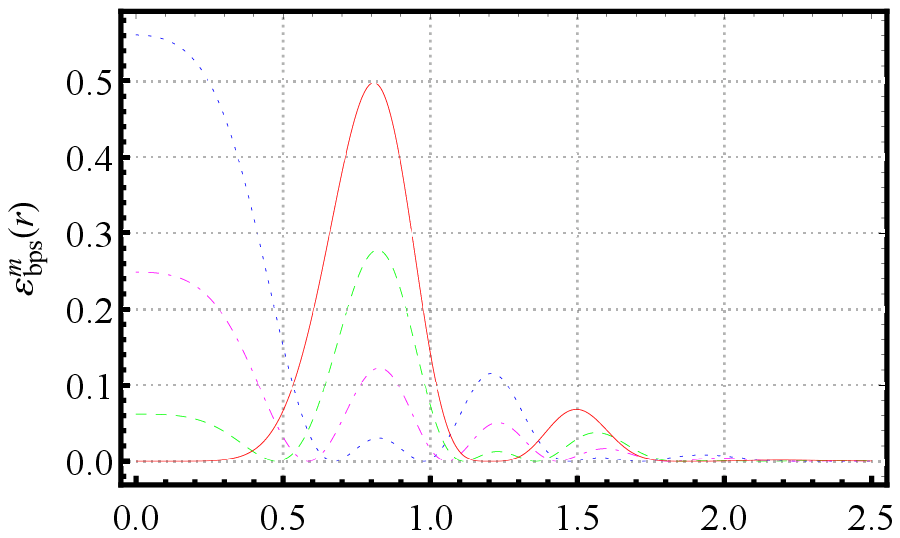}\vspace{0.5cm} %
\includegraphics[width=8.4cm]{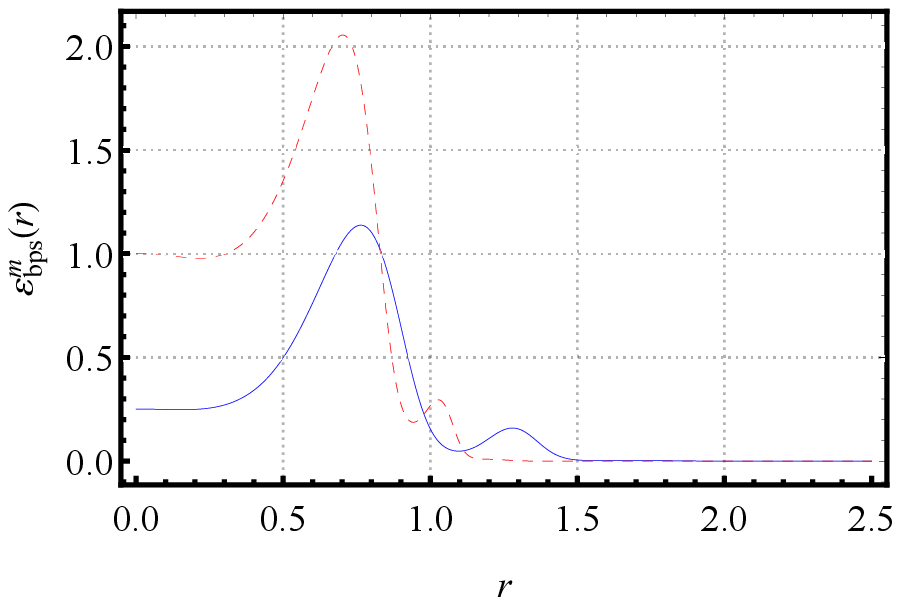}
\caption{Numerical solutions to the BPS magnetic energy density $\protect%
\varepsilon _{bps}^{M}\left( r\right) $ obtained via {Eqs}. (\protect
\ref{xx3}) and (\protect\ref{xx4}) for different values of $\protect\alpha $%
. Conventions as in the Fig. 9.}
\label{fig_8}
\end{figure}

On the other hand, to study the effects caused by the different values of $%
\alpha $ on the shape of $\varepsilon _{bps}^{M}$, we calculate%
\begin{eqnarray}
\frac{d}{dr}\varepsilon _{bps}^{M} &=&4h^{3}\left[ \alpha -\cos ^{2}\left(
3\pi h\right) \right]  \notag \\[0.08in]
&&\hspace{-1cm}\times \left[ \frac{{}}{{}}[3\pi h\sin \left( 3\pi h\right)
-\cos \left( 3\pi h\right) ]\cos \left( 3\pi h\right) +\alpha \right] \frac{%
dh}{dr}\text{,}
\end{eqnarray}%
{via which one gets that} the condition $(\varepsilon
_{bps}^{M})^{\prime }=0$ gives rise to%
\begin{eqnarray}
&\cos ^{2}\left( 3\pi h\right) =\alpha \text{,}&  \label{eq1x} \\[0.2cm]
&\left[ 3\pi h\sin \left( 3\pi h\right) -\cos \left( 3\pi h\right) \right]
\cos \left( 3\pi h\right) =-\alpha \text{,}&  \label{eq1y}
\end{eqnarray}%
where the first one is the same Eq. (\ref{eq1}).

The value of the magnetic energy density at $r=0$ can be verified to be%
\begin{equation}
\varepsilon _{bps,0}^{M}=\varepsilon _{bps}^{M}\left( r=0\right) =\left(
\alpha -1\right) ^{2}\text{,}  \label{med00}
\end{equation}%
i.e. the same result obtained in the previous case, see the Eq. (\ref{med0}%
). Here, we have again used $h\left( r=0\right) =1$.

From this point on, we implement the same prescription used previously, i.e.
we solve {Eqs.} (\ref{eq1x}) and (\ref{eq1y}) for a particular $%
\alpha $, from which we get the corresponding roots $h_{i}$'s. We then
calculate $\varepsilon _{bps}^{M}$ via Eq. (\ref{med1}) at those different $%
h_{i}$'s and categorize the resulting values as local maxima or minima of
the solution, from which we interpret the magnetic energy density according
to its profile and describe how its dimensions depend on $\alpha $.

In what follows, we again consider only {some} ranges of values for $%
\alpha $, for the sake of illustration. 
\begin{table}[t]
\caption{Results for $\protect\alpha =0.25$. The exact roots of Eq. (\protect
\ref{eq1x}) (first column), the approximate roots of Eq. (\protect\ref{eq1y}%
) (second column) and the values of $\protect\varepsilon _{bps}^{M}\left(
r\right) $ calculated at these last roots (third column). The magnetic
energy density at the origin is given by $\protect\varepsilon %
_{bps,0}^{M}=0.5625$.}
\label{tab_4}%
\begin{ruledtabular}
		\begin{tabular}{clddc}
			{}&Roots of (\protect\ref{eq1x})&
			\multicolumn{1}{r}{\textrm{Roots of (\protect
\ref{eq1y})}}&{}&{Values of $\varepsilon _{bps}^{M}$}\\
			\colrule
			& & & & \\ [-0.25cm]
			{}&$8/9$& 0.8367 & &{0.0304} \\		
			{}&$7/9$& 0.6790 & &{0.1153} \\
			{}&$5/9$& 0.5055 & & {0.0040} \\
			{}&$4/9$& 0.3563 & &{0.0080} \\
			{}&$2/9$& 0.1812 & &{$6\times 10^{-5}$} \\
			{}&$1/9$& 0.0736 & &{$3\times 10^{-6}$}
		\end{tabular}
	\end{ruledtabular} 
\end{table}

\paragraph{The case $\protect\alpha =0$:}

It mimics the {behavior} already identified for the magnetic field
itself, i.e. in this case, the Eq. (\ref{eq1x}) provides the roots given by 
{Eqs.} (\ref{H3}), (\ref{H2}) and (\ref{H1}). As explained
previously, at these points, $B(r)$ vanishes identically, so as the magnetic
energy density, from which we conclude that the resulting solution to $%
\varepsilon _{bps}^{M}\left( r\right) $ presents therefore $3$ (three)
different zeros.

In addition, Eq. (\ref{eq1y}) can be reduced to%
\begin{equation}
\left[ 3\pi h\sin \left( 3\pi h\right) -\cos \left( 3\pi h\right) \right]
\cos \left( 3\pi h\right) =0\text{,}
\end{equation}%
which is satisfied by $\cos \left( 3\pi h\right) =0$ (i.e. the Eq. (\ref%
{eq1x}) for $\alpha =0$) and%
\begin{equation}
3\pi h\sin \left( 3\pi h\right) -\cos \left( 3\pi h\right) =0\text{,}
\label{texx}
\end{equation}%
which stands for a transcendental expression whose roots must be determined
numerically.

Once that we have obtained the roots of\ $\cos \left( 3\pi h\right) =0$ and
clarified that they represent the zeros of $\varepsilon _{bps}^{M}$, we now
focus our attention on the solutions of Eq. (\ref{texx}) above and their
meaning concerning the magnetic energy profile. 
\begin{table}[b]
\caption{Results for $\protect\alpha =0.50$. Conventions as in the Table IV.
The magnetic energy density at the origin is given by $\protect\varepsilon %
_{bps,0}^{M}=0.25$.}
\label{tab_5}%
\begin{ruledtabular}
		\begin{tabular}{clddc}
			{}&Roots of (\protect\ref{eq1x})&
			\multicolumn{1}{r}{\textrm{Roots of (\protect
\ref{eq1y})}}&{}&{Values of $\varepsilon _{bps}^{M}$}\\
			\colrule
			& & & & \\ [-0.25cm]
			{}&$11/12$& 0.8400 & &{0.1225} \\		
			{}&$9/12$& 0.6749 & &{0.0506} \\
			{}&$7/12$& 0.5109 & & {0.0163} \\
			{}&$5/12$& 0.3490 & &{0.0034} \\
			{}&$3/12$& 0.1933 & &{$0.0003$} \\
			{}&$1/12$& 0.0571 & &{$6\times 10^{-7}$}
		\end{tabular}
	\end{ruledtabular} 
\end{table}

The point is that the transcendental Eq. (\ref{texx}) admits three numerical
roots given approximately\ by%
\begin{eqnarray}
h_{1} &=&0.6830\text{,} \\[0.2cm]
h_{2} &=&0.3635\text{,} \\
h_{3} &=&0.0913\text{,}
\end{eqnarray}%
{from which one calculates the values }$\varepsilon
_{bps,1}^{M}=0.2075${, }$\varepsilon _{bps,2}^{M}=0.0148${\
and }$\varepsilon _{bps,3}^{M}=0.00001${, respectively,} which stand
for local maxima of the magnetic energy density. The solution to $%
\varepsilon _{bps}^{M}$ obtained for $\alpha =0$ therefore presents three
different peaks for intermediate values of $r$.

Moreover, when $\alpha =0$, the value of $\varepsilon _{bps}^{M}$ at $r=0$
can be verified to be%
\begin{equation}
\varepsilon _{bps,0}^{M}=1\text{,}
\end{equation}%
according to the previous Eq. (\ref{med00}).

In summary, the results above reveal that the magnetic energy's profile
starts from its global maximum at $r=0$ and then interpolates between three
zeros and three peaks before vanishing in the asymptotic limit, {the
resulting solution forming a lump surrounded by three rings.}

\paragraph{The case $0<\protect\alpha <1$:}

In this case, Eq. (\ref{eq1x}) admits the $6$ (six) analytical roots given
by Eq. (\ref{hi}), from which one gets that $\varepsilon _{bps}^{M} $ now
possess six different zeros, this way mimicking the behaviour of $B(r)$
itself.

In addition, the approximate roots which come from the numerical study of
Eq. (\ref{eq1y}) give rise to those peaks {(rings)} which define the
internal structure inherent to the magnetic energy's profile, with the value
of such energy density at $r=0$ {(i.e. the magnitude of the centered
lump)} still being given by Eq. (\ref{med00}).

Again for the sake of illustration, we solve the transcendental Eq. (\ref%
{eq1y}) for $\alpha =0.25$, $\alpha =0.50$ and $\alpha =0.75$, from which we
display the numerical results in the second columns of the {tables
IV, V and VI}, respectively. These tables also bring the exact roots of Eq. (%
\ref{eq1x}) (i.e. the points at which $\varepsilon _{bps}^{M}$ vanishes, see
the first column) and the corresponding values of magnetic energy's peaks
(third column).

The comparison between the results displayed in {the tables IV, V and
VI} reveals that the solution to $\varepsilon _{bps}^{M}$ for $0<\alpha <1$
has a nonvanishing value at $r=0$ which only eventually represents a global
maximum of the corresponding profile. In addition, driven by the magnetic
field itself, the solution to $\varepsilon _{bps}^{M}$ presents an internal
structure with six zeros and six peaks, {i.e. it stands for a lump
now surrounded by six rings.}

The results also show that both the zeros and the peaks {(i.e. the
position of the rings)} of the magnetic energy density behave in the same
way as those of $B(r)$ itself, i.e. as $\alpha $ increases, the first, third
and fifth (second, fourth and sixth) zeros and peaks of $\varepsilon
_{bps}^{M}$ move towards (outwards) the origin. 
\begin{table}[t]
\caption{Results for $\protect\alpha =0.75$. Conventions as in the Table IV.
The magnetic energy density at the origin is given by $\protect\varepsilon %
_{bps,0}^{M}=0.0625$.}
\label{tab_6}%
\begin{ruledtabular}
		\begin{tabular}{clddc}
			{}&Roots of (\protect\ref{eq1x})&
			\multicolumn{1}{r}{\textrm{Roots of (\protect
\ref{eq1y})}}&{}&{Values of $\varepsilon _{bps}^{M}$}\\
			\colrule
			& & & & \\ [-0.25cm]
			{}&$17/18$& 0.8433 & &{0.2778} \\		
			{}&$13/18$& 0.6708 & &{0.0125} \\
			{}&$11/18$& 0.5161 & & {0.0375} \\
			{}&$7/18$& 0.3414 & &{0.0008} \\
			{}&$5/18$& 0.2044 & &{$0.0007$} \\
			{}&$1/18$& 0.0388 & &{$3\times 10^{-8}$}
		\end{tabular}
	\end{ruledtabular} 
\end{table}

\paragraph{The case $\protect\alpha >1$:}

In addition, the solutions for $\alpha =1.5$ and $\alpha =2$ possess an
internal structure which {interpolates} between different minima
which are not equal to zero {(multiple volcano configurations)}. As
before, an increasing $\alpha $ continuously reduces the number of roots
which solve Eq. (\ref{eq1y}). This number finally vanishes for $\alpha $
large enough. This extreme case (i.e. the absence of roots) gives rise to a
magnetic energy distribution whose numerical curve forms a lump centered at $%
r=0$ with no internal structure (i.e. it vanishes monotonically in limit $%
r\rightarrow \infty $).


\section{Summary and perspectives\label{sec2 copy(1)}}

{We have investigated BPS solitons inherent to a gauged baby
Skyrme scenario {in which} a nontrivial function ({which plays}
the role of the magnetic permeability of the medium) multiplies the Maxwell
term. {This enlarged} model possesses a well-defined BPS structure 
{which allows us }to attain the self-dual equations and {a}
lower bound for the {total} energy {which is} proportional to
the topological charge, {as expected}. The convenient choice of 
{the} magnetic permeability function allows the construction of
magnetic fields whose profiles dramatically differ from their standard
counterparts.}

{We have focused our attention on those time-independent
configurations with radial symmetry. This way, we have divided our
investigation into two different branches according to the expression for
the magnetic permeability (which depends on the parameter $\alpha \in 
\mathbb{R}$}). {We have solved the corresponding BPS equations numerically
by means of a finite-difference scheme, from which we have observed {%
that} the profiles of the BPS solutions dramatically depend on $\alpha $.} 
{We} have then explained analytically the formation of internal
structures {based on} the values of $\alpha $. {In the
sequence,} we have identified the effects caused by different values of $%
\alpha $ on the general shape of the numerical solutions, including the
emergence of a sophisticated structure which interpolates between different
peaks {(rings)} and zeros of the BPS magnetic sector. We have also
studied how the dimensions which distinguish such structures (such as
amplitude and radius) depend on $\alpha $ itself.

Next, we have depicted the numerical solutions for different values of $%
\alpha $ and fixed values of the other constants of the model. The figures
not only confirm our analytical predictions concerning the occurrence of
internal structures and their dimensions, but they also allow us to observe
how these structures can be extremely intricate depending on $\alpha $
itself. {{In this sense, we have we have pointed out that,%
} when $\alpha $ is relatively small, the magnetic sector develops an
structure {formed by multiple concentric rings positioned at}
intermediate values of the radial coordinate. On the other hand, it is
important to highlight that $\alpha $ sufficiently large leads to numerical
solutions which mimic the canonical shape (differing from their standard
counterparts only by their dimensions, such as amplitude and radius), 
{i.e.} {a single lump with a compact profile} centered at $r=0$
}{and which therefore vanishes in the asymptotics.}

The values of both the magnetic field and the magnetic energy density at the
origin depend on $\alpha $ explicitly and this fact helps us to define how
the corresponding profiles must be classified according to the position of
their global maximums. {Moreover,} the lower bound for the total
energy does not depend on $\alpha $, being exactly the same for all the
cases considered in the present manuscript.

We are currently studying the existence of BPS skyrmions in the context of a
theory in which the dynamics of the gauge field is controlled by the
    Born-Infeld action now enlarged to include a nontrivial {magnetic
permeability}, from which we expect the obtainment of BPS solutions with
internal structures which can be thought as {generalisations} of the
ones presented in this {work.} The results will be reported in a
future contribution.

\begin{acknowledgments}
The authors thank Prof. Dion\'{\i}sio Bazeia and Prof. Lukasz Stepien for motivating discussions. This work was financed in part by the Coordena\c{c}\~{a}o
de Aperfei\c{c}oamento de Pessoal de N\'{\i}vel Superior - Brasil (CAPES) -
Finance Code 001, the Conselho Nacional de Pesquisa e Desenvolvimento Cient%
\'{\i}fico e Tecnol\'{o}gico - CNPq and the Funda\c{c}\~{a}o de Amparo \`{a}
Pesquisa e ao Desenvolvimento Cient\'{\i}fico e Tecnol\'{o}gico do Maranh%
\~{a}o - FAPEMA (Brazilian agencies). In particular, {J. A. and A. C. S.} thank the
full support from CAPES (via a PhD scholarship and a Postdoctoral fellowship,
respectively), {R. C.} acknowledges the support from the grants CNPq/306724/2019-7, FAPEMA/Universal-01131/17 and FAPEMA/Universal-00812/19, and E. H. thanks the support from the grant CNPq/309604/2020-6.
\end{acknowledgments}

\end{document}